\documentclass[11pt]{article}
\usepackage{amsmath,amssymb}
\usepackage{hyperref}
\usepackage{graphicx}
\usepackage{epstopdf}
\newcommand{\bea}{\begin{eqnarray}}
\newcommand{\eea}{\end{eqnarray}}

\newcommand{\ft}[2]{{\textstyle\frac{#1}{#2}}}

\newcommand{\spa}{\quad, \quad}

\renewcommand{\Im}{\operatorname{Im}}
\renewcommand{\Re}{\operatorname{Re}}

   \newcommand\CN{{\cal{N}}}
    \newcommand\CS{{\cal{S}}}
\newcommand\CR{{\cal{R}}}

\newsavebox{\uuunit}
\sbox{\uuunit}
    {\setlength{\unitlength}{0.825em}
     \begin{picture}(0.6,0.7)
        \thinlines
        \put(0,0){\line(1,0){0.5}}
        \put(0.15,0){\line(0,1){0.7}}
        \put(0.35,0){\line(0,1){0.8}}
       \multiput(0.3,0.8)(-0.04,-0.02){10}{\rule{0.5pt}{0.5pt}}
     \end {picture}}

\numberwithin{equation}{section}
\begin{document}

\thispagestyle{empty}

\hfill LMU-ASC 29/11\\[-3ex]

\hfill MPP-2011-91 \\
\vskip 6mm

\begin{center}
{\bf\LARGE
Nernst branes in gauged supergravity}

\vspace{10mm}

{\large
{\bf Susanne Barisch$^{*,\times}$,}
{\bf Gabriel Lopes Cardoso$^+$,}
{\bf Michael Haack$^*$,}\\ \vskip 3mm
{\bf Suresh Nampuri$^*$,}
{\bf Niels A. Obers$^\dagger$}
\vspace{1cm}

{\it $^*$
 Arnold Sommerfeld Center for Theoretical Physics \\ [1mm]
Ludwig-Maximilians-Universit\"at M\"unchen \\ [1mm] 
Theresienstrasse 37, 80333 M\"unchen, Germany \\ [5mm] 
\it $^\times$
Max-Planck-Institut f\"ur Physik \\ [1mm]
F\"ohringer Ring 6, 80805 M\"unchen, Germany \\ [5mm] 
$^+$
CAMGSD, Departamento de Matem\'atica\\ [1mm]
Instituto Superior T\'ecnico, Universidade T\'ecnica de Lisboa\\ [1mm]
Av. Rovisco Pais, 1049-001 Lisboa, Portugal \\ [5mm] 
$^\dagger$ 
The Niels Bohr Institute \\ [1mm]
Blegdamsvej 17, 2100 Copenhagen {\O}, Denmark 
}}

\end{center}
\vspace{8mm}

\begin{center}
{\bf ABSTRACT}\\
\end{center}

We study static black brane solutions in the context of $N = 2$ $U(1)$
gauged supergravity in four dimensions. Using the formalism of
first-order flow equations, we construct novel extremal black brane
solutions including examples of Nernst branes, i.e.\ extremal black
brane solutions with vanishing entropy density. We also discuss
a class of non-extremal generalizations which is captured by the first-order
formalism.

\clearpage
\setcounter{page}{1}

\tableofcontents

\section{Introduction}

Black holes play an important role as testing grounds for theories of
quantum gravity. They originally arise as solutions of the equations
of classical gravitation which encapsulate regions of space-time where
the curvature becomes of the order of Planck length and classical
gravitation breaks down. Thus, a quantum theory of gravity must be
able to make statements about these systems. Further, in classical
gravitation, there is a correspondence between physical observables
that characterize the horizon of the black hole, such as its surface
area and the surface gravity, and thermodynamic quantities such as
entropy and temperature, and once this correspondence is made, the
laws of black hole mechanics can be rewritten as the laws of
thermodynamics of black holes.
Therefore, an essential requirement for any quantum theory of gravity
is that it has to be able to derive this correspondence from first
principles and more generally, derive the thermodynamics of black
holes from a microscopic statistical physical viewpoint. Hence, black
hole thermodynamics and in particular questions of microstate counting
of black hole entropy have been an active area of research in string
theory.

Amongst these intriguing systems exists a subclass of black holes
which pose extremely interesting problems that have been the target of
concentrated research in string theory. These are a subset of charged
black holes called extremal black holes. Charged black holes  arise as
solutions of gravity coupled to gauge fields and in general can carry
electric and magnetic charge quantum numbers w.r.t.\ these gauge
fields. Extremal black holes are charged black holes that carry the
minimum mass possible in the theory for a given set of charge quantum
numbers, and their mass is uniquely fixed in a given theory in terms
of the charges
and the asymptotic values of the scalar fields.
These are zero temperature black holes which radically violate the
third law of thermodynamics -- the so called Nernst Law,\footnote{The
original Nernst heat theorem states that entropy changes go to zero at
zero temperature. The strong form of the law which states that entropy
goes to zero at zero temperature was formulated by Max Planck, for
pure crystalline homogenous materials \cite{Planck}.} which states that the entropy
of a thermodynamic system vanishes at zero temperature. Because these
black holes gravitationally attract matter, they reduce the entropy of
the region of space-time outside the horizon, and hence for
consistency with the second law of thermodynamics, they must have a
corresponding increase in entropy. Hence these black holes have a very
high non-zero entropy at zero temperature. Some of the major successes
in the study of the statistical mechanics of black holes in string
theory has been for this class of black holes. In particular, their entropy can be 
given a statistical interpretation \cite{Strominger:1996sh,Maldacena:1997de} and exact
formulae for the microscopic degeneracies of extremal black holes are
known for certain classes of
supersymmetric extremal black holes. These include black holes in ${N}\,=\,4 \,,\,
D\,=\,4$ string theory or ${N}\,=\,4 \,,\, D\,=\,5$ string theory (see \cite{Dijkgraaf:1996it,
Dijkgraaf:1996xw,Shih:2005uc, {Jatkar:2005bh}, {Dabholkar:2008zy}}), and in ${N}\,=\,8 \,,\,
D\,=\,4$ string theory (see \cite{Sen:2008ta}).

Simultaneously, there is another critical way in which black holes
play an essential role in string theory. String theory naturally
generates a beautiful and extremely powerful correspondence between
the dynamics of fields in gravitational backgrounds that are
asymptotically AdS and conformal field theories on the boundary of this
AdS space. The AdS/CFT correspondence implies the equality of the
Hilbert spaces of the bulk theory and the field theory at the
boundary, whose operator content is defined by the boundary values of
the bulk fields.  Hence every state in the bulk has a corresponding
state in the field theory. Black holes which are states in the bulk
are represented as thermal ensembles in the dual field theory at the
same temperature as the black hole. The dynamics of bulk fields in the
black hole background therefore provides information on the
interaction of the corresponding operators in the thermal
ensemble of the dual field theory. Since the AdS/CFT correspondence is
a strong-weak coupling duality, non-perturbative aspects of the boundary
field theory can, in principle, be understood in terms of
perturbative calculations in gravity and vice versa.\footnote{Incidentally, the AdS/CFT correspondence
also lies at the heart of the entropy computation for extremal black holes in 
string theory: The leading order entropy of 
extremal black holes with a local $AdS_3$ factor in the near horizon geometry (arising from combining the 
usual $AdS_2$ factor with an internal circle direction)
can be obtained by relating their Bekenstein-Hawking entropy to the Cardy-Hardy-Ramanujan
formula for the leading term in the entropy of a 2D CFT (see 
\cite{Strominger:1997eq,Balasubramanian:1998ee,Dabholkar:2006tb}).}

A lot of effort has been put in applying this
paradigm to extract information about field theory systems from a gravitational
perspective, recently specifically about those field theories 
that could underlie condensed matter physics, (for a comprehensive introduction to the various aspects of this field
see the reviews \cite{Hartnoll:2009sz,Herzog:2009xv,McGreevy:2009xe,Hartnoll:2009qx,Horowitz:2010gk,Sachdev:2010ch}). 
One of the most interesting and rich problems in such
systems is the problem of understanding phase transitions. In some important cases these are  
inaccessible to perturbative calculations on the field
theory side and as such offer a ready application ground to applying
the principles of AdS/CFT or, more generally, in applying knowledge of
gravitational physics to the corresponding field theory dual using
gauge-gravity duality. Black holes play a vital role in this new
application.

Amongst the most challenging classes of phase transitions are quantum
critical phase transitions which are phase transitions that occur in
condensed matter systems at zero temperature and which are driven by
quantum fluctuations. In order to analyze these transitions
gravitationally, one must necessarily focus on the gravitational duals
of zero temperature thermal ensembles in field theory -- the extremal
black holes. Further, if one expects to find results applicable to
real condensed matter systems, these black holes must obey the usual
thermodynamic properties of condensed matter systems.
All extremal black holes obey the first two laws of thermodynamics and
we therefore need to explore extremal black holes that do not violate
the third law. Hence finding these 'Nernst black holes' is a relevant
problem with potentially rich implications. Smooth Nernst configurations in AdS have already been 
found in \cite{Gauntlett:2009dn,Horowitz:2009ij,D'Hoker:2009bc}.

Once we find a class of Nernst black holes, we need to have a systematic way of
classifying these objects. Phase transitions in field theory are
classified by their universality classes and distinguishing between
various kinds of Nernst black holes could help in identifying those
relevant for a particular field theory. One way that suggests itself as a useful approach in classification is
in terms of a behavior that is uniquely exhibited by extremal
configurations -- namely the attractor mechanism. This mechanism was discovered 
when studying uncharged scalar fields in an extremal black
hole background.\footnote{The dynamics of these fields is encoded in a
low-energy effective action
arising from string theory.} In the absence of fluxes,
the attractor mechanism forces the values of the fields on the horizon
to be independent of their asymptotic values and fixes them solely in
terms of the charges of the black hole. Hence the extremal black hole
horizon serves as a fixed point in moduli space (see \cite{Ferrara:1995ih,{Strominger:1996kf}, {hep-th/9602136}, {hep-th/9603090},Ferrara:1997tw,hep-th/0507096}).\footnote{In general there 
can also be flat directions in moduli space. However, the entropy of an extremal black hole
does not depend on the values of the moduli
associated with these flat directions, so that it still makes sense to view the extremal black hole
horizon as a fixed point in moduli space.}
Therefore, classifying extremal black holes can be restated as
classifying these fixed points in moduli space and identifying any
non-universality in the fixed point flow behavior.
\par
The purpose of this paper is therefore to make a step towards addressing
the analogous problem in the presence of fluxes by
studying static (black) brane solutions in the context of $N = 2 \,\, U(1)$
gauged supergravity in four dimensions with only vector multiplets. Related work on extending the attractor mechanism 
to the case with gaugings can be found in \cite{Bellucci:2008cb,Goldstein:2009cv,Cacciatori:2009iz,Goldstein:2010aw,Dall'Agata:2010gj,Hristov:2010ri,Kachru:2011ps}.
Our study starts with obtaining first-order flow equations for
extremal black branes, building on and extending recent work done in
\cite{Cacciatori:2009iz,Dall'Agata:2010gj,Hristov:2010ri}.
Using this, we first study a number of new exact (non-Nernst)
solutions
both in the single scalar and STU model. These include, in particular,
generalizations of the (extremal limit) of a solution recently discussed in the context of AdS/CMT in
\cite{Charmousis:2010zz}, as well as a solution that
interpolates between $AdS_4$ and $AdS_2 \times \mathbb{R}^2$.
We then show how our formalism can be used in the STU model to find
explicit examples
of extremal black brane solutions in four dimensions with vanishing
entropy, which we denote as {\it Nernst branes}.
We also show how the first-order formalism can be extended to
non-extremal solutions and we give examples thereof.

The paper is organized as follows. In section
\ref{sec:first-order-rew}, we first study the equations of motion
governing the flow of the fields towards a fixed point,
in the presence of fluxes. The presence of a fixed point in moduli
space basically halves the degrees of freedom and hence the flow equations for
the fields can be formulated as first-order ODEs. In four-dimensional ungauged
supergravity these equations were
first constructed and solved for black holes  in \cite{Ferrara:1997tw, Denef:2000nb}.
In the case of black holes in ungauged supergravity, the curvature of
the horizon is positive and inversely proportional to the square
of the spherical radius of the horizon. A vanishing entropy simply
corresponds to a vanishing horizon size and a curvature larger than
Planck scale curvature, which clearly pushes these objects outside the
regime of analysis of supergravity. In general, for systems with a
non-vanishing horizon curvature, classical supergravity analysis
applies when the radius of curvature is macroscopic w.r.t.\ both the Planck as
well as the string scales. In four dimensions, both these scales are related by the string
coupling constant as $l_p\,=\,g_s\,l_s$. At the
horizon, if the string coupling is fixed and the horizon does not
encode information about the quantum numbers of the internal states of
the system, that is if it has vanishing entropy, the string coupling
value at the fixed point is a universal number independent of charges
and independent of its asymptotic value. For 'small' black holes with
vanishing entropy, the string coupling constant is
generally zero or infinite \cite{Sen:1995in} and in either case, for non-vanishing value
of the dimensionfull curvature, the dimensionless curvature will blow
up in either string or Planck units. Hence, we should naturally turn
towards black objects with Ricci-flat horizons. These are black branes
in string theory. In the absence of fluxes, in an asymptotically flat
space, black objects in $D=4$ can have only a spherical topology. Hence, we
need to turn on fluxes to generate black branes in non-flat
backgrounds. 
\par
In the presence of fluxes, extremal supersymmetric configurations in
four dimensions were first discovered by \cite{Cacciatori:2009iz}
and subsequently discussed in \cite{Dall'Agata:2010gj,Hristov:2010ri}. 
The first-order formalism in the presence of fluxes was first
presented in \cite{Dall'Agata:2010gj}. 
There are significant differences in this formalism vis-a-vis the
ungauged case. The two most crucial ones being that there is a
symplectic constraint that restricts the choices of charges given the
fluxes and secondly, the phase of the
effective `central' charge (which we denote by $\gamma$) is no longer
constant as in the ungauged case, but is a dynamical quantity with an
equation of its own. The first-order formalism simply depends on the
existence of fixed points in moduli space, and is independent of
whether the solutions being considered are supersymmetric or
non-supersymmetric. Hence, after re-deriving the supersymmetric first-order equations 
of \cite{Dall'Agata:2010gj}, as a generalization, we indicate how to develop a first-order
equation for moduli flows in a non-supersymmetric black brane
background, and we give examples thereof. We
demonstrate how the invariance of the action under certain
transformation operations on the fluxes can lead to alternative
first-order rewritings of the differential equations of flow. We
finish this formalism section by identifying the necessary conditions 
for the near-horizon geometry to be $AdS_2 \times \mathbb{R}^2$.
\par
In section \ref{sec:exact-sol}, we begin to explore the solution space
of the first-order flow equations. We first choose the simplest
prepotential encoding one complex vector multiplet scalar field, given
by $F\,= \,-\,i\,X^0\,X^1$, and  obtain a class of  general fixed
point solutions representing non-Nernst black branes. These turn out
to be generalizations of the solution discussed in
\cite{Charmousis:2010zz}. This serves as a consolidating check on the
formalism. This solution, like an ungauged solution, has a constant
phase $\gamma$. As another example of a solution with constant $\gamma$,
we show a full numerical solution that
interpolates between a near-horizon $AdS_2 \times \mathbb{R}^2$ geometry and an asymptotic $AdS_4$
geometry (for the prepotential $F\,= \,-\,(X^1)^3/X^0$). 
We then proceed to explore the possibility to have non-constant $\gamma$. 
So far we have only found local geometries whose curvature and string 
coupling become large at some point. 
It might be possible, though, that they describe the asymptotic region of a global solution 
once higher derivative corrections are taken into account. It would be interesting to
further investigate the existence of well behaved solutions with non-constant $\gamma$,
given that such solutions would be radically different
from the constant phase sector and have no
counterpart in the ungauged case. 
\par
In section \ref{sec:nernst},
we finally pursue the question of finding  Nernst brane solutions. For
this purpose, we explore axion-free solutions of the fixed point flow
equations in the STU model, and write down a solution with constant
$\gamma$ and a fixed point at the zero of the radial
coordinate. At this point, the scalar that parametrizes the dilaton in the heterotic frame
 flows to zero. Moreover, at this point, the metric has a
coordinate singularity and the area density in the constant time and
constant radial coordinate hyperplane vanishes. The geometry near this
fixed point has an infinitely long radial throat which suppresses
fluctuations in the scalar fields such that their solutions become 
independent of their asymptotic values. We will take this infinite throat property to 
mean that the solution is extremal, and the vanishing of the area density to mean that the 
solution has zero entropy density. However, these
solutions asymptote to geometries that are not $AdS_4$ and, thus, it is not clear 
which role they might play in the gauge/gravity correspondence. 
Nevertheless, they are well behaved in that the curvature invariants are finite.
We end this section with a few comments on this Nernst brane solution.
\par
In section \ref{sec:deform}, we present a generalization of the
first-order formalism to non-extremal black branes and show how in
certain cases, the moduli flow in these backgrounds can be encoded in
first-order equations.
\par

%%%%%%%%%%%%%%%%%%%%%%%%%%%%%%%%%%%%%%%%%%%%%%%%%%%%%%

\section{First-order flow equations for extremal black branes \label{sec:first-order-rew}}

In the following, we will be interested in extremal black brane solutions of $N=2$ $U(1)$ gauged
supergravity in four dimensions with vector multiplets. 
First-order flow equations for supersymmetric black holes and black branes
were recently obtained in \cite{Dall'Agata:2010gj} by a rewriting of the action, 
where they were given in terms of physical scalar fields $z^i=Y^i/Y^0$ ($i = 1, \dots, n$).
Here, we will re-derive them by working in big moduli space, so that the resulting
first-order flow equations will now be expressed in terms of the $Y^I$. Also, we will not restrict ourselves to supersymmetric
solutions only. The formulation in big moduli space becomes particularly useful when discussing
the coupling to higher-derivative curvature terms \cite{deWit:1996ag}.

\subsection{Flow equations in big moduli space}

Following \cite{Cacciatori:2009iz,Dall'Agata:2010gj,Hristov:2010ri}
we make the ansatz for the black brane line element,
\begin{equation}
ds^2 = - {\rm e}^{2 U} \, dt^2 + {\rm e}^{-2 U} \left( dr^2 + {\rm e}^{2 \psi} (dx^2 + dy^2) \right) \;,
\label{eq:line-blackb}
\end{equation}
where $U = U(r) \,,\, \psi = \psi (r)$.  The black brane will be supported by scalar fields that
only depend on $r$.

The Lagrangian we will consider is given in \eqref{eq:lagN2}. It is written in terms of fields $X^I$ of big moduli
space.  The rewriting of this Lagrangian as a sum of squares of first-order flow equations will, however,
not be in terms of the $X^I$, but rather in terms of rescaled variables $Y^I$ defined by
\begin{equation}
Y^I = {\rm e}^{A} \, {\tilde X}^I = {\rm e}^{A} \, {\bar \varphi} \, X^I \;.
\end{equation}
Here $A = A(r)$ denotes a real factor that will be determined to be given by
\begin{equation}
A = \psi - U \;,
\label{eq:A-psi-U}
\end{equation}
while $\bar \varphi$ denotes a phase with a
$U(1)$-weight that is opposite to the one of $X^I$.  Thus, the $\tilde X^I = \bar \varphi X^I$  
denote homogeneous coordinates that are $U(1)$ invariant and satisfy \eqref{eq:constraint-sugra} (with $X^I$ replaced by  $\tilde X^I$) as well as
\begin{equation}
N_{IJ} \, {\cal D}_r X^I \, {\cal D}_r {\bar X}^J = N_{IJ} \, \tilde{X}'^I \bar{\tilde{X}}'^J 
\;,
\label{eq:kin-scalars}
\end{equation} 
where $\tilde{X}'^I = \partial_r \tilde{X}^I$.
Observe that in view of \eqref{eq:constraint-sugra},
\begin{equation}
{\rm e}^{2 A} = - N_{IJ} \, Y^I \, {\bar Y}^J \;,
\label{eq:A-Y}
\end{equation}
and that
\begin{equation}
{\rm e}^{2 A} \, A' = - \tfrac12 \, N_{IJ} \, \left( Y'^I \, {\bar Y}^J + Y^I \, {\bar Y}'^J \right) \;,
\label{eq:relAp-Y}
\end{equation}
where we used the second homogeneity equation of 
\eqref{eq:homog-rel}.

We will first discuss electrically charged extremal black branes in the presence of electric fluxes $h_I$ only, so that for the time being the flux potential \eqref{eq:flux-pot} reads
\begin{equation}
V(\tilde{X}, \bar{\tilde{X}}) = \left( N^{IJ}   - 2 \, {\tilde X}^I \, {\bar {\tilde X}{}^J }\right) \, h_I \, h_J \;.
\label{eq:pot-flux-elec}
\end{equation}
Subsequently, we will extend
the first-order rewriting to the case of dyonic charges as well as dyonic fluxes.

We take $F_{tr}^I = E^I (r)$ as well as
$X^I = X^I (r)$.
Inserting the line element \eqref{eq:line-blackb} into the action \eqref{eq:lagN2} yields the one-dimensional
Lagrangian 
\begin{equation} \label{1dlagrange}
{\cal L}_{1d} = \sqrt{-g} \, L - Q_I \, E^I \;,
\end{equation}
where $L$ is given in \eqref{eq:lagN2} and the $Q_I$ denote the electric charges.
Extremizing with respect to $E^I$ yields
\begin{equation}
- {\rm e}^{2 \psi - 2 U} \, {\rm Im} {\cal N}_{IJ} \, E^J = Q_I \;,
\end{equation}
and hence
\begin{equation} \label{EQ}
E^I = - {\rm e}^{2 U - 2 \psi} \left[\left( {\rm Im} {\cal N}\right)^{-1}\right]^{IJ} \, Q_J \;.
\end{equation}
The associated one-dimensional action reads, 
\begin{eqnarray} \label{s1d}
- S_{1d} &=& \int dr \, {\rm e}^{2 \psi} \left\{U'^2 - \psi'^2 + 
N_{IJ} \, {\tilde X}'^I \, \bar{\tilde X}'^J
- \ft12 \, {\rm e}^{2 U - 4 \psi} \, Q_I \left[ \left({\rm Im} {\cal N}\right)^{-1}\right]^{IJ}
\, Q_J \right. \nonumber\\
&& \qquad \qquad \qquad \left.   +  g^2 \, {\rm e}^{-2U} V(\tilde{X}, \bar{\tilde{X}})  \right\} \nonumber\\
 && + \int dr \frac{d}{dr} \left[ {\rm e}^{2 \psi} \left( 2 \psi' - U' \right) \right] \;,
\label{eq:action-1d}
\end{eqnarray}
in accordance with \cite{Dall'Agata:2010gj} for the case of black branes. 
Next, we rewrite \eqref{s1d} in terms of the rescaled variables $Y^I$.  We also find it convenient to introduce
the combination
\begin{equation}
q_I = {\rm e}^{U - 2 \psi + i \gamma} \left( Q_I - i \, g \, {\rm e}^{ 2 (\psi - U )} h_I
\right) \;,
\label{eq:rel-q-Q-h}
\end{equation}
where $\gamma$ denotes a phase which can depend on $r$.
Using 
\begin{equation}
{\tilde X}'^I 
= {\rm e}^{-A} \, \left(Y'^I - A' \, Y^I \right) \;, 
\end{equation}
as well as \eqref{eq:relAp-Y}, we obtain the intermediate result
\begin{eqnarray}
- S_{1d} &=& \int dr \, {\rm e}^{2 \psi} \Big\{U'^2 - \psi'^2   \nonumber\\
&& \left.  + 
{\rm e}^{-2 A} \, N_{IJ} \, \left(Y'^I - {\rm e}^{ A} \, N^{IK} \, {\bar q}_K\right)
 \, \left({\bar Y}'^J - {\rm e}^{ A} \, N^{JL} \, q_L\right)
+ \left( A' +  \Re \Big[ {\tilde X}^I \, q_I \Big]\right)^2
\right.  \nonumber\\
&&  - \ft12 \, {\rm e}^{2 U - 4 \psi} \, Q_I \left[ \left({\rm Im} {\cal N}\right)^{-1}\right]^{IJ}
\, Q_J -   q_I \, N^{IJ} \, {\bar q}_J  -  \left(
 \Re \left[{\tilde X}^I \, q_I \right]\right)^2
 +  g^2 \, {\rm e}^{-2U} V(\tilde{X}, \bar{\tilde{X}})   \Big\} \nonumber\\
&& + 2 \int dr  \,  {\rm e}^{2 \psi} \, \Re \left[ 
{\tilde X}'^I  \, q_I \right]
 + \int dr \frac{d}{dr} \left[ {\rm e}^{2 \psi} \left( 2 \psi' - U' \right) \right] \;.
\label{eq:action-1d-interm2}
\end{eqnarray}
Next, using the identity,  
\begin{equation}
 - \tfrac12 \, \left[\left({\rm Im} {\cal N}\right)^{-1}\right]{}^{IJ}
= 
N^{IJ} + 
{\tilde X}^I \, {\bar {\tilde X}}{}^J + {\tilde X}^J \, {\bar {\tilde X}}{}^I \;,
\end{equation}
as well as the explicit form of the potential \eqref{eq:pot-flux-elec}, we obtain
\begin{eqnarray}
- S_{1d} &=& \int dr \, {\rm e}^{2 \psi} \Big\{U'^2 - \psi'^2  \nonumber\\
&& \left.  + 
{\rm e}^{-2 A} \, N_{IJ} \,  \left(Y'^I - {\rm e}^{ A} \, N^{IK} \, {\bar q}_K\right)
\,  \left({\bar Y}'^J - {\rm e}^{ A} \, N^{JL} \, q_L\right) 
+ \left( A' +  \Re \Big[{\tilde X}^I \, q_I \Big] \right)^2
\right.  \nonumber\\
&&
+ 2 \, {\rm e}^{2 U - 4 \psi} \, Q_I {\tilde X}^I \, Q_J \bar{\tilde X}^J 
- \left( \Re \left[{\tilde X}^I \, q_I \right]\right)^2
 - 2 \,  g^2 \, {\rm e}^{-2U} \, h_I \, {\tilde X}^I \, h_J \, \bar{\tilde{X}}{}^J 
  \Big\} \nonumber\\
 &&+ 2 \int dr  \,  {\rm e}^{2 \psi} \, \Re \left[ 
{\tilde X}'^I  \, q_I \right]
 + \int dr \frac{d}{dr} \left[ {\rm e}^{2 \psi} \left( 2 \psi' - U' \right) \right] \;.
  \label{eq:action-1d-interm4}
\end{eqnarray}
Inserting the expression \eqref{eq:rel-q-Q-h}
into the fourth line of \eqref{eq:action-1d-interm4}
yields
\begin{eqnarray}
  2 \int dr  \,  {\rm e}^{2 \psi} \, \Re \left[ {\tilde X}'^I \, q_I \right]
&=&  2 \int dr  \, \frac{d}{dr} \left[ {\rm e}^U \, \Re \left( {\rm e}^{i \gamma} \, {\tilde X}^I \, Q_I   \right) 
+ g \, {\rm e}^{2 \psi - U} \, \Im \left({\rm e}^{i \gamma} \, {\tilde X}^I \, h_I \right) 
\right]
\nonumber\\
&& - 2 \,  \int dr \, {\rm e}^{U} \, U' \, \Re \left[ {\rm e}^{i \gamma} \,{\tilde X}^I \, Q_I \right] \nonumber\\
&& - 2 \,g \,  \int dr \, {\rm e}^{2 \psi - U} \, \left(2 \psi' - U' \right) \, \Im \left[ 
{\rm e}^{i \gamma} \,{\tilde X}^I \, h_I \right]  \\
&& + 2 \,  \int dr \,\gamma'   \left[ {\rm e}^U \Im \left( {\rm e}^{i \gamma} \,{\tilde X}^I \, Q_I \right) - g \,  {\rm e}^{2 \psi - U} \Re \left( 
{\rm e}^{i \gamma} \,{\tilde X}^I \, h_I \right)  \right] \nonumber \;.
\end{eqnarray}
Combining the terms proportional to $\psi'^2$ and to $\psi'$ into a perfect square, and the terms proportional
to $U'^2$ and to $U'$ into a perfect square, yields
\begin{eqnarray}
- S_{1d} = S_{\rm BPS} + S_{\rm TD} \;,
\end{eqnarray}
where
\begin{eqnarray}
\label{eq:action-BPS}
S_{\rm BPS} &=& 
\int dr \, {\rm e}^{2 \psi} \left\{\Big[U' - {\rm e}^{U - 2 \psi}
\Re \left( {\rm e}^{i \gamma} \, {\tilde X}^I \, Q_I   \right) 
+ g \, {\rm e}^{- U} \, \Im \left({\rm e}^{i \gamma} \, {\tilde X}^I \, h_I \right) 
\Big]^2 \right. \nonumber\\
&& \left. - \left(\psi' + 2 \, g \, {\rm e}^{-U} \, \Im \left[{\rm e}^{i \gamma} \,
{\tilde X}^I \, h_I \right] \right)^2 \right. \\
&& \left.  + 
{\rm e}^{-2 A} \, N_{IJ} \,  \left(Y'^I - {\rm e}^{ A} \, N^{IK} \, {\bar q}_K\right)
 \, \left({\bar Y}'^J - {\rm e}^{ A} \, N^{JL} \, q_L\right) 
+ \left( A' +  \Re \Big[{\tilde X}^I \, q_I \Big] \right)^2 + \Delta \right\} \;, \nonumber
\end{eqnarray}
and
\begin{eqnarray}
\Delta 
&=& 2 \left[   {\rm e}^{U - 2 \psi} \, \Im \left({\rm e}^{i \gamma} \,  {\tilde X}^I \, Q_I \right)
- g \, {\rm e}^{-U} \Re \left({\rm e}^{i \gamma} \, {\tilde X}^I \, h_I \right) \right] \nonumber\\
&& \quad \left[  \gamma' +  {\rm e}^{U - 2 \psi} \, \Im \left({\rm e}^{i \gamma} \,  {\tilde X}^I \, Q_I \right)
+ g \, {\rm e}^{-U} \Re \left({\rm e}^{i \gamma} \, {\tilde X}^I \, h_I \right) \right]
 \;.
\end{eqnarray}
Finally,
\begin{eqnarray}
\label{eq:action-TD}
S_{\rm TD} &=&  \int dr  \, \frac{d}{dr} \left[ 
 {\rm e}^{2 \psi} \left( 2 \psi' - U' \right) +
2 \, 
{\rm e}^U \, \Re \left({\rm e}^{i \gamma} \,   {\tilde X}^I \, Q_I   \right) 
+ 2 \, g \, {\rm e}^{2 \psi - U} \, \Im \left( {\rm e}^{i \gamma} \, {\tilde X}^I \, h_I \right)
\right] \;. \nonumber\\
\end{eqnarray}
Setting the squares in $S_{\rm BPS}$ to zero gives
\begin{eqnarray}
U' &=& {\rm e}^{U - 2 \psi}
\Re \left( {\rm e}^{i \gamma} \, {\tilde X}^I \, Q_I   \right) 
- g \, {\rm e}^{- U} \, \Im \left({\rm e}^{i \gamma} \, {\tilde X}^I \, h_I \right) \;,\nonumber\\
\psi' &=& - 2 \, g \, {\rm e}^{-U} \, \Im \left[{\rm e}^{i \gamma} \, {\tilde X}^I \, h_I \right]\;, \nonumber\\
A' &=& - \Re \left[{\tilde X}^I \, q_I \right] \;, \nonumber\\
Y'^I &=& {\rm e}^{ A} \, N^{IK} \, {\bar q}_K \;,
\label{eq:flow-electric}
\end{eqnarray}
while demanding the variation of $\Delta$ to be zero yields
\begin{eqnarray}
 {\rm e}^{U - 2 \psi} \, \Im \left({\rm e}^{i \gamma} \,  {\tilde X}^I \, Q_I \right)
- g \, {\rm e}^{-U} \Re \left({\rm e}^{i \gamma} \, {\tilde X}^I \, h_I \right) = 0 
\label{eq:constraint-Q-h}
\end{eqnarray}
as well as
\begin{eqnarray}
 \gamma'  = -{\rm e}^{U - 2 \psi} \, \Im \left({\rm e}^{i \gamma} \,  {\tilde X}^I \, Q_I \right)
- g \, {\rm e}^{-U} \Re \left({\rm e}^{i \gamma} \, {\tilde X}^I \, h_I \right)\; .
\end{eqnarray}
Note that the first-order flow equations for the $Y^I$ and for $A$ 
are consistent with one another: the latter is a consequence of the former by virtue of \eqref{eq:relAp-Y}. The flow equations given above are obtained by varying \eqref{eq:action-BPS} with respect to the various
fields and setting the individual terms to zero. Therefore, the solutions to these flow equations
describe a subclass, but certainly not the entire class of solutions to the equations of motion stemming from the one-dimensional action \eqref{s1d}.

Comparing the flow equations \eqref{eq:flow-electric} with the ones obtained in the supersymmetric
context in \cite{Dall'Agata:2010gj} shows that the flow equations derived above are the ones for supersymmetric
black branes, and that the phase $\gamma$ is to be identified with the phase $\alpha$ of \cite{Dall'Agata:2010gj}
via $\gamma' = - \left( \alpha' + {\cal A}_r \right)$, with ${\cal A}_r$ given in 
\eqref{eq:kah-conn}.

Next, we study the dyonic case, with charges $(Q_I, P^I)$ and fluxes $(h_I, h^I)$ turned on.
The above results can be easily extended by first writing the term $Q ({\rm Im} {\cal N})^{-1} Q$
in the action \eqref{eq:action-1d} as
\begin{eqnarray}
V_{\rm BH} &=& - \tfrac12 \, Q_I \, \left[({\rm Im} {\cal N})^{-1}\right]^{IJ} \, Q_J 
\nonumber\\
&=& 
\left( N^{IJ} + 2 \, \tilde{X}^I \, \bar{\tilde{X}}^J \right) Q_I \, Q_J \nonumber\\
&=& g^{i \bar j} \, {\cal D}_i Z \, \bar{\cal D}_{\bar j} {\bar Z} + |Z|^2 \;,
\end{eqnarray}
where we used \eqref{eq:N-id-X} to write $V_{\rm BH}$ in terms of $Z = - Q_I \, X^I$. Turning on magnetic charges amounts to extending $Z$ to \cite{hep-th/9602136}
\begin{eqnarray}
Z = P^I \, F_I - Q_I \, X^I = \left(P^I \, F_{IJ} - Q_J \right) X^J = - \hat{Q}_I \, X^I \;,
\end{eqnarray}
where 
\begin{eqnarray}
\hat{Q}_I = Q_I - F_{IJ} \, P^J \;.
\end{eqnarray}
Similarly, the flux potential with dyonic fluxes can be obtained from the one with purely electric fluxes by the replacement of $h_I$ by 
\begin{eqnarray}
\hat{h}_I = h_I - F_{IJ} \, h^J \;,
\end{eqnarray}
cf.\ \eqref{eq:flux-pot}. 

Thus, formally the action looks identical to before, and we can adapt the computation given above to the case of dyonic charges and fluxes by replacing $Q_I$ and $h_I$ with ${\hat Q}_I$ and ${\hat h}_I$. Performing these replacements in \eqref{eq:rel-q-Q-h} as well yields
\begin{equation}
q_I = {\rm e}^{U - 2 \psi + i \gamma} \left( \hat{Q}_I - i \, g \, {\rm e}^{ 2 (\psi - U )} \hat{h}_I
\right) \;.
\label{eq:q-Qhat-hhat}
\end{equation}
The above procedure results in 
\begin{equation}
- S_{1d} = S_{\rm BPS} + S_{\rm TD} + S_{\rm sympl} \;,
\label{eq:action-sympl}
\end{equation}
where $S_{\rm BPS}$ and $S_{\rm TD}$ are given as in \eqref{eq:action-BPS} and \eqref{eq:action-TD}, respectively,
with $Q_I$ and $h_I$ replaced by $\hat{Q}_I$ and $\hat{h}_I$, and with $q_I$ now given by
\eqref{eq:q-Qhat-hhat}. The third contribution, $ S_{\rm sympl}$,
is given by
\begin{equation} 
 S_{\rm sympl} =  g \, \int dr \left(Q_I \, h^I - P^I \, h_I \right) \;.
 \label{eq:sympl-explicit}
 \end{equation}
Observe that this term is constant, independent of the fields, and hence it does not
contribute to the variation of the fields.  Imposing the 
constraint ${\rm S}_{1d} =0$ (which is the Hamiltonian constraint, to be discussed below) on a solution
yields the condition
\begin{equation}
Q_I \, h^I - P^I \, h_I = 0\;,
\label{eq:sympl-constr}
\end{equation}
in agreement with \cite{Dall'Agata:2010gj} for the case of black branes.  The condition \eqref{eq:sympl-constr}
can also be written as 
\begin{equation}
{\rm Im}\left( \hat Q_I \, N^{IJ} \, \bar{\hat{h}}_J \right) = 0\ .
\label{eq:symplsimpl}
\end{equation}

The flow equations 
are now given by
\begin{eqnarray}
U' &=& {\rm e}^{U - 2 \psi} \, 
 \Re \left[{\rm e}^{i \gamma} \, 
 {\tilde X}^I \, \hat{Q}_I \right] - g \, {\rm e}^{-U} \, \Im \left[{\rm e}^{i \gamma} \,
 {\tilde X}^I \, \hat{h}_I \right]
  \;, \nonumber\\
\psi' &=& - 2 \, g \, {\rm e}^{-U} \, \Im \left[{\rm e}^{i \gamma} \, {\tilde X}^I \, \hat{h}_I \right]\;, \nonumber\\
A' &=& - \Re \left[{\tilde X}^I \, q_I \right] \;,  \nonumber\\
Y'^I &=& {\rm e}^A \, N^{IJ} \, \bar{q}_J \;, \nonumber\\
\gamma' &=& {\rm e}^{U-2 \psi} {\rm Im} \left( {\rm e}^{i \gamma} \, {\tilde Z} \right) + g \,
{\rm e}^{-U} {\rm Re} \left( {\rm e}^{i \gamma} \, {\tilde W} \right) \;,
\label{eq:flow-U-psi-all}
\end{eqnarray}
where $\tilde{Z}$ and $\tilde{W}$ denote $Z$ and $W$ with $X$ replaced by $\tilde{X}$, as in 
\eqref{eq:tilde-W}.
Inspection of the flow equations \eqref{eq:flow-U-psi-all} yields
\begin{equation}
A' = \left(\psi - U \right)' \;,
\end{equation}
and hence we obtain \eqref{eq:A-psi-U} (without loss of generality).  

Observe that, as before, the flow equations
for $A$ and $Y^I$ are consistent with one another. The latter can be recast into
\begin{eqnarray}
\begin{pmatrix} (Y^I - {\bar Y}^I)' \\ (F_I - {\bar F}_I)' 
\end{pmatrix} = 
- 2 i  \, {\rm e}^{-\psi } \, \Im
\begin{pmatrix} {\rm e}^{i \gamma} \,
N^{IK} \, {\hat Q}_K \\ {\rm e}^{i \gamma} \,{\bar F}_{IK} \, N^{KJ} \, {\hat Q}_J 
\end{pmatrix}
 + 2 i \, g \, {\rm e}^{\psi - 2 U} \, \Re
\begin{pmatrix}
{\rm e}^{i \gamma} \, N^{IK} \, {\hat h}_K \\ {\rm e}^{i \gamma} \, {\bar F}_{IK} \, N^{KJ} \, {\hat h}_J 
\end{pmatrix} ,
\label{eq:attrac-Q-h}
\end{eqnarray}
where here $F_I = \partial F(Y)/\partial Y^I$. 
Each of the vectors appearing in this expression transforms as a symplectic vector under $\rm Sp(2(n+1))$transformations, i.e. as
\begin{equation}
\begin{pmatrix} Y^I  \\ F_I 
\end{pmatrix} \rightarrow 
\begin{pmatrix} U^I{}_J  & Z^{IJ}
\\
W_{IJ}  & V_I{}^J
\end{pmatrix} 
\begin{pmatrix} Y^J  \\ F_J 
\end{pmatrix} \;,
\label{eq:sympl-trafo}
\end{equation}
where $U^T \, V - W^T \, Z = \cal{I}$.  For instance, under symplectic transformations,
\begin{equation}
N^{IJ} \, \bar{\hat h}_J \rightarrow {\cal S}^I{}_K \, N^{KL} \, \bar{\hat h}_L \;\;\;,\;\;\;
F_{IJ} \rightarrow \left(V_I{}^L \, F_{LK} + W_{IK} \right) [{\cal S}^{-1}]^K{}_J \;,
\end{equation}
where ${\cal S}^I{}_J = U^I{}_J + Z^{IK} \, F_{KJ}$ \cite{de Wit:1996ix}.  Using this, it can be easily checked that the last vector
in \eqref{eq:attrac-Q-h} transforms as in \eqref{eq:sympl-trafo}.

The constraint \eqref{eq:constraint-Q-h} becomes
\begin{eqnarray}
 {\rm e}^{U - 2 \psi} \, \Im \left({\rm e}^{i \gamma} \, Z(Y) \right)
- g \, {\rm e}^{-U} \Re \left({\rm e}^{i \gamma} \, W(Y) \right) = 0\; ,
\label{eq:cond-Z-W-Y}
\end{eqnarray}
where
\begin{eqnarray}
Z(Y) &=& P^I \, F_I (Y) - Q_I \, Y^I \;, \nonumber\\
W(Y) &=& h^I \, F_I (Y) - h_I \, Y^I \;.
\label{eq:Z-W-Y}
\end{eqnarray}
Using the flow equations \eqref{eq:flow-U-psi-all}, we find that 
the constraint \eqref{eq:cond-Z-W-Y} is equivalent to the condition
\begin{equation}
q_I \, Y^I = {\bar q}_I \, {\bar Y}^I \;.
\label{eq:q-Y-real}
\end{equation}

The phase $\gamma$ is not an independent degree of freedom.  It can be expressed in terms of 
$Z(Y)$ and $W(Y)$ as \cite{Dall'Agata:2010gj}
\begin{equation}
{\rm e}^{-2 i \gamma} = \frac{Z(Y) - i g \, {\rm e}^{2(\psi-U)} \, W(Y)}{
\bar{Z}(\bar{Y}) + i g \, {\rm e}^{2(\psi-U)} \, \bar{W}(\bar{Y})} \;.
\label{eq:gamma-Z-W}
\end{equation}
When $\gamma = k \pi$ ($k \in \mathbb{Z}$), equation \eqref{eq:gamma-Z-W} implies
\begin{equation}
Z(Y) = {\bar Z}(\bar Y) \;\;\;,\;\;\; W(Y) = - {\bar W} (\bar Y) \;.
\label{eq:Z-bZ}
\end{equation}

Differentiating \eqref{eq:cond-Z-W-Y}
with respect to $r$ yields a flow equation for $\gamma'$ that has to be consistent
with the last flow equation in \eqref{eq:flow-U-psi-all}.  We proceed to check consistency of these two flow equations.
Multiplying \eqref{eq:cond-Z-W-Y} with $\exp{(2 \psi -U)}$ and differentiating the resulting expression with
respect to $r$ yields
\begin{equation}
\Im \Big[ i \, q_I \, Y^I 
\left( \gamma' - 2 g \, {\rm e}^{-\psi} \, {\rm e}^{i \gamma}  \, W(Y) \right) \Big] =0 \;,
\end{equation}
where we used the flow equations for $Y^I$ and for $(\psi - U)$.  Using \eqref{eq:q-Y-real}, this results in 
\begin{equation}
 \gamma' = 2 g \, {\rm e}^{-\psi} \, \Re \Big[{\rm e}^{i \gamma} \, W(Y) \Big]\; ,
 \label{eq:flow-g-W}
\end{equation}
which, upon using \eqref{eq:cond-Z-W-Y}, equals the flow equation for $\gamma$ given in \eqref{eq:flow-U-psi-all}.
Thus, we conclude that upon imposing the constraint \eqref{eq:cond-Z-W-Y}, the flow equation for $\gamma'$
given in \eqref{eq:flow-U-psi-all} is automatically satisfied.

Summarizing, we obtain the following independent flow equations.  Since ${\rm e}^{2A}$ is expressed in terms 
of $Y^I$, it is not an independent quantity, cf.\ \eqref{eq:A-Y}.  Since $U = \psi - A$ on a solution, $U$ is also not
an independent quantity.  The independent flow equations are thus
\begin{eqnarray}
\psi' &=&  2 \, g \, {\rm e}^{-\psi} 
\, \Im \left[{\rm e}^{i \gamma} \, W(Y) \right]\;, \nonumber\\
Y'^I &=& {\rm e}^{\psi -U} \, N^{IK} \, {\bar q}_K \;,
\label{eq:flow-independent}
\end{eqnarray}
with $q_K$ given in \eqref{eq:q-Qhat-hhat} and with $\gamma$ given in \eqref{eq:gamma-Z-W} (observe
that the latter is, in general, $r$-dependent).
In addition, a solution to these flow equations has to satisfy the reality condition \eqref{eq:q-Y-real}
as well as the symplectic constraint \eqref{eq:sympl-constr}.
The flow equations for the scalar fields $Y^I$ can equivalently be written in the form
\eqref{eq:attrac-Q-h}. Observe that in view of \eqref{eq:A-Y} and 
\eqref{eq:q-Y-real}, the number of 
independent variables in the set $(U, \psi, Y^I)$ is the same as in the set $(U, \psi, z^i=Y^i/Y^0)$, which was used in  \cite{Dall'Agata:2010gj}. Indeed, the flow equations \eqref{eq:flow-independent} are equivalent to the ones presented there and, thus, they describe supersymmetric brane solutions. 

Observe that the right hand side of the flow equations \eqref{eq:flow-independent}
may be expressed in terms of the $Y^I$ only by redefining the radial variable into
$\partial/\partial \tau = {\rm e}^{\psi} \, \partial/\partial{r}$, in which case they become
\begin{eqnarray}
\frac{\partial \psi}{\partial \tau} &=&  2 \, g
\, \Im \left[{\rm e}^{i \gamma} \, W(Y) \right]\;, \nonumber\\
\frac{\partial Y^I}{\partial \tau} &=& {\rm e}^{ -i \gamma}  \,
 N^{IK} \, 
\left( \bar{\hat{Q}}_K + i \, g \, {\rm e}^{ 2 A} \bar{\hat{h}}_K\right) \;,
\label{eq:flow-independent-tau}
\end{eqnarray}
with $A$ expressed in terms of the $Y^I$ according to \eqref{eq:A-Y}.

Let us now discuss the Hamiltonian constraint mentioned above.
For a Lagrangian density $\sqrt{-g}\, (\,\frac12 \mathfrak{R}\,+\,{\cal L}_M\,)\,$ it is given by the variation of the action w.r.t.\ $g^{00}$ as 
\begin{equation} \label{hamilton}
\tfrac12  \mathfrak{R}_{00}\,+\,\frac{\delta{\cal L}_M}{\delta g^{00}}\,-\tfrac12 \,g_{00}\,\Big(\tfrac12  \mathfrak{R}\,+\,{\cal L}_M
\Big)\,=\,0\ .
\end{equation}
Using the matter Lagrangian \eqref{eq:lagN2} as well as the metric ansatz \eqref{eq:line-blackb},
and replacing the gauge fields by their charges, as in \eqref{EQ},  gives
\begin{eqnarray}
 {\rm e}^{2\psi} \left\{ U'^2\,-\,\psi'^2 \,+\, N_{IJ}\,{\tilde X}^I\;\!\! ' \,\bar{\tilde X}^J\; \!\! '
+\,g^2\, {\rm e}^{-2\,U\,}\,V_{\rm tot} (\tilde{X}, \bar{\tilde{X}})\, \right\} 
+\,\Big[\,{\rm e}^{2\psi}\,(2\psi'\,-\,2\,U'\,)\Big]' = 0\;,
\label{eq:hamilton-s1d}
\end{eqnarray}
where $V_{\rm tot}$ denotes the combined potential
\begin{eqnarray}
V_{\rm tot} (\tilde{X}, \bar{\tilde{X}}) &=& g^2 \, V(\tilde{X}, \bar{\tilde{X}}) + {\rm e}^{-4 A} \,
V_{\rm BH} (\tilde{X}, \bar{\tilde{X}}) \nonumber\\
&=&  g^2 \, \left[ N^{IJ} \, \partial_I \tilde{W} \partial_{\bar J} \bar{\tilde{W}} -2 |\tilde{W}|^2 \right]
+ {\rm e}^{-4A} \, \left[ N^{IJ} \, \partial_I \tilde{Z} \partial_{\bar J} {\bar{\tilde{Z}}} +2 |\tilde{Z}|^2 \right] \;.
\label{eq:pot-tot}
\end{eqnarray}
The Hamiltonian constraint \eqref{eq:hamilton-s1d} can now be rewritten (up to a total derivative term) as
\begin{eqnarray}
L_{\rm BPS} + L_{\rm sympl} = 0 \;,
\label{eq:outc-ham}
\end{eqnarray}
where $L_{\rm BPS}$ and $L_{\rm sympl}$ denote the integrands of \eqref{eq:action-BPS} 
(with the charges and fluxes replaced by their hatted counterparts) and \eqref{eq:sympl-explicit},
respectively. Since $L_{\rm BPS} =0$ on a solution to the flow equations, it follows that
the Hamiltonian constraint reduces to the symplectic constraint \eqref{eq:sympl-constr}.  The total derivative
term vanishes by virtue of the flow equation for $A= \psi - U$.

In the following, we briefly check that \eqref{eq:attrac-Q-h} reproduces the standard flow equations
for supersymmetric domain wall solutions in $AdS_4$.
Setting $Q_I = P^I = 0$, and choosing the phase $\gamma = \pi/2$,
we obtain $\psi = 2 U \,,\, A = U$ as well as
\begin{eqnarray}
\begin{pmatrix} (Y^I - {\bar Y}^I)' \\ (F_I - {\bar F}_I)' 
\end{pmatrix} = 
i \,g \, 
\begin{pmatrix} h^I \\ h_I
\end{pmatrix} \;,
\label{eq:attr-dw}
\end{eqnarray}
which describes the supersymmetric domain wall solution of \cite{Behrndt:2001mx}. Observe that the choice 
$\gamma = 0$ leads to a solution of the type \eqref{eq:attr-dw} with the imaginary part of $(Y^I, F_I)'$
replaced by the real part.
The flow equations \eqref{eq:attr-dw} can be easily integrated
and yield
\begin{eqnarray}
\begin{pmatrix} Y^I - {\bar Y}^I \\ F_I - {\bar F}_I
\end{pmatrix} = 
 i \, g 
\begin{pmatrix} H^I \\ H_I
\end{pmatrix} =  
 i \, g 
\begin{pmatrix} \alpha^I + h^I \, r \\ \beta_I + h_I \, r
\end{pmatrix} \;,
\label{eq:attr-dw-int}
\end{eqnarray}
where $(\alpha_I, \beta^I)$ denote integration constants. This solution satisfies
$W(Y) = {\bar W}(\bar Y)$, which is precisely the condition \eqref{eq:cond-Z-W-Y}.
In addition, using \eqref{eq:A-Y}, we obtain
${\rm e}^{2U} = g \left(H^I \, F_I(Y) - H_I \, Y^I \right)$.

%%%%%%%%%%%%%%%%%%%%%%%%%%%%%%%%%%%%%%%%%%%%%%%%%%%%%%

\subsection{Generalizations of the first-order rewriting
\label{sec:gen-first-order}}

In the preceding section the first order rewriting was done for a general prepotential. In mi\-ni\-mal gauged supergravity other first order rewritings are possible. For a detailed discussion see appendix \ref{mingauge}. 
Moreover, in the first-order rewriting performed above, what was used
operationally was the fact that
i) for an arbitrary charge and flux configuration the 1-D bulk
Lagrangian can be written  as a sum of squares and
ii) imposing the symplectic constraint \eqref{eq:sympl-constr}, the vanishing of the squares
gives a vanishing bulk Lagrangian, which in
turn is equal to the Hamiltonian density.
This ensures that both the equations of motion as well as the
Hamiltonian constraint are satisfied.
Given this procedure, we may identify invariances of the action under
transformations of the charges and/or fluxes, since
any such transformation will automatically produce a new, possibly
physically distinct, rewriting of the action.

In the ungauged case in four dimensions this has been already explored
in terms of an $\CS$-matrix
that operates on the charges, while keeping the action invariant
\cite{Ceresole:2007wx}.\footnote{See also \cite{Lopes Cardoso:2007ky} for related
transformations in five dimensions.}
In the gauged case, the presence of both charges and fluxes allows for a wider
class of transformations, in which both sets of quantum numbers are
transformed. Thus finding transformations
on both charges and fluxes that leave the action invariant allows for
more general rewritings. Here we illustrate
this with two possible types of transformations, one based on charge
transformations and the other on fluxes. It is in principle
also possible to find combinations thereof, but we do not attempt to do so explicitly.

\subsubsection{Charge transformations}

The combined potential \eqref{eq:pot-tot} arising in the 1-D action
consists of a sum of two positive definite
terms, one associated with charges and the other one with fluxes.  The part
arising from charges can be written as 
\begin{equation}
\label{eq:BHpot}
V_{\rm BH} = -\tfrac{1}{2}  {\cal{Q}}^T  {\cal{M}} {\cal{Q}} \;,
\end{equation}
where ${\cal{Q}} = (P^I, Q_I)$ and the symplectic matrix ${\cal{M}}$
takes the form (following the notation of \cite{Ceresole:2007wx})
\begin{equation}
\label{Mmatrix}
{\cal{M}} = {\cal{I}} M  \spa M =  \left( \begin{matrix} D & C \\ B  &
A  \end{matrix} \right) \spa {\cal{I}} =
\left( \begin{matrix} 0 & -  \mathbb{I}  \\   \mathbb{I} & 0
\end{matrix} \right)
\end{equation}
with
\begin{equation}
A = - D^T = \Re \CN (\Im \CN)^{-1} \spa C = (\Im \CN)^{-1} \spa B = -
\Im \CN - \Re \CN (\Im \CN)^{-1} \Re \CN \;.
\end{equation}
Note that $M^2 =- \mathbb{I}$.
Given a matrix $\CS$ that transforms the charge vector ${\cal{Q}}$ to
${\cal{Q}}' = \CS{\cal{Q}}$ and obeying
$\CS^T {\cal{M}} \CS = {\cal{M}} $ the charge potential remains
invariant. Moreover, if in addition we
choose a set of fluxes ${\cal{H}}=(h^I,h_I)$ such that their
symplectic product \eqref{eq:sympl-constr} with the charges remains
invariant,
\begin{equation}
\label{sympinv}
{\cal{H}}^T {\cal{I}} {\cal{Q}} = {\cal{H}}^T  {\cal{I}} \CS {\cal{Q}} \;,
\end{equation}
then the 1-D action \eqref{eq:outc-ham} remains invariant.
Consequently, the same action can have two different rewritings, one
being the usual supersymmetric rewriting in terms of the original
charges and fluxes and the other being a rewriting in terms of
the new '$\CS$-transformed' charges and old fluxes.  Since the latter
rewriting is distinct from the original supersymmetric one, it would
necessarily be non-supersymmetric.

In further detail, if one furthermore assumes that $\CS$ is
symplectic, it was shown in \cite{Ceresole:2007wx}
that the invariance of \eqref{eq:BHpot} implies that it needs to satisfy
\begin{equation}
\label{SMcom}
[\CS, M] =0
\end{equation}
with $M$ defined in \eqref{Mmatrix}. In the present case with gauging,
one needs to require in addition \eqref{sympinv}.
A sufficient condition for this is that ${\cal{H}} = \CS \, {\cal{H}}$ so
that $\CS$ leaves the flux vector unchanged \cite{Ceresole:2007wx}.
This shows that one can generically satisfy this condition by choosing
a matrix $\CS$ that only mixes non-zero charges
(in the original configuration) and by only turning on fluxes that are dual
to those charges that are set to zero.

Examples of the $\CS$-matrix and charge and flux configurations
satisfying the conditions stated above
will be given in section \ref{sec:exnonsusy} below. Explicit rewritings
for a constant $\CS$-matrix in the ungauged case were
given in  \cite{Ceresole:2007wx} for the 4-D case and in  \cite{Lopes
Cardoso:2007ky} for 5-D. In particular, in the 4-D case
a constant S-matrix implies that one can define a new ``fake
superpotential'' that gives rise to the same black hole
potential as $Z$.
It must be noted that though in principle one can construct
non-constant $\CS$-matrices that are field dependent and which satisfy
these criteria,  for the purpose of first-order rewriting, it is
non-trivial to employ these matrices since their derivatives then
appear in the action so that the squaring technique in the action
becomes highly involved.
We will only give explicit examples of the equations in the case of a
constant $\CS$-matrix. We note that the $\cal S$-matrix can be viewed
as a solution generating technique for generating non-supersymmetric solutions out of
supersymmetric solutions to the equations of motion of the 1-D action \eqref{1dlagrange}.

\subsubsection{Flux transformations}

There are two distinct types of flux transformations that leave the
action invariant and can hence be used to generate
new solutions.

The first one is in spirit analogous to the charge transformation
discussed above. One may write the flux potential \eqref{eq:flux-pot}
as
\begin{equation}
V_g = {\cal{H}}^T \, {\cal{L}} \, {\cal{H}}
\end{equation}
with ${\cal{H}} = (h^I, h_I)$ and the matrix ${\cal{L}}$ given by
\begin{equation}
\label{Lmatrix}
{\cal{L}} = \left( \begin{matrix} F T \bar F   & -   F T  \\  - T
\bar F   & T    \end{matrix} \right)
\end{equation}
in terms of  $F_{IJ}$  and the matrix
\begin{equation}
T^{IJ} = N^{IJ} - 2 X^I \bar{X}^J\ .
\end{equation}
Note that we thus have $(FT)_I{}^J = F_{IK} T^{KJ}$, $(T \bar F)^I{}_J
= T^{IK} \bar F_{KJ}$, $(FT \bar F)_{IJ} = F_{IK} T^{KL} \bar F_{LJ}$.
We now look for transformations on the fluxes
${\cal{H}} \rightarrow {\cal{H}}' = \CR {\cal{H}}$ that leave the flux
potential invariant, i.e.
$\CR^T {\cal{L}} \CR = {\cal{L}}$ as well as the symplectic constraint
${\cal{H}}^T {\cal{I}} {\cal{Q}} = {\cal{H}}^T \CR^T {\cal{I}}
{\cal{Q}}$. This is a new feature of the gauged case, examples of
which will be given in section \ref{sec:exnonsusy}.
Note that if one finds an $\CS$-matrix transforming the charges and an
$\CR$-matrix transforming
the fluxes, that simultaneously keep the symplectic constraint fixed,
it is possible to perform a non-supersymmetric
rewriting with both transformed charges and fluxes.

A second possibility, which is more model-dependent, occurs if there
is a flux configuration  $\hat{h}_*$ that
contributes  vanishingly to the flux potential $V_g\,=\,
\hat{h}_I\,(\,N^{IJ}\,-\,2\,X^I\,\bar{X}^J\,)\,
\bar{\hat{h}}_J$, cf.\ \cite{Galli:2010mg}.\footnote{We thank Stefanos Katmadas for pointing this out to us.} 
If, in addition, the charges that are symplectically dual to the fluxes
in $\hat{h}_*$  are not turned on, then this flux vector becomes a
flat direction in flux space. This means that for a given 1-D action
one can introduce these starred fluxes by adding a null term to the
action as well as the Hamiltonian, while not affecting the symplectic
constraint. Then a new rewriting of the action can be implemented with
the old charges but the fluxes changed by the addition of the
corresponding $\hat{h}_*$.  An example of this arises in the STU model
where a candidate $\hat{h}_*$ is simply the flux configuration with
only $h_0$ turned on. One can then easily show that 
\begin{equation}
\left(h + h_* \right)_I \, T^{IJ} \left(h + h_* \right)_J = 
h_I \, T^{IJ} h_J \;.
\end{equation}
Hence the replacement $h \rightarrow h+h_*$ can be made in the action, provided the charge $P^0$
is not turned on, and a new rewriting can be
achieved. Note that this type of rewriting satisfies the supersymmetry
equations and is hence also supersymmetric. Note also that higher
derivative corrections will almost certainly lift this apparent
degeneracy in solution space.
Finally, this new rewriting technique could be combined with the two
previous ones for appropriate charge and flux configurations to
achieve further rewritings.

%%%%%%%%%%%%%%%%%%%%%%%%%%%%

\subsubsection{Non-supersymmetric examples \label{sec:exnonsusy}}

As an illustration of the transformations discussed above, we now give
several examples in the two particular models
for which we later in sections \ref{sec:exact-sol} and \ref{sec:nernst}
obtain new extremal black brane solutions. The transformations below
(along with other examples one might wish to construct) can thus be
used to generate non-supersymmetric extremal black brane solutions
from these.

We start with the $F(X)=-i X^0 X^1$ model.  In this case one can
compute a constant (symplectic) $\CS$-matrix
satisfying \eqref{SMcom} (see  eq.\ (4.8) of
\cite{Ceresole:2007wx}),
\begin{equation}
\CS = \left( \begin{matrix} -\cos \theta \sigma_3 & -i \sin \theta \sigma_2 \\
i \sin \theta \sigma_2 &  - \cos \theta \sigma_3   \end{matrix} \right) \;,
\end{equation}
with $\sigma_a$ the standard Pauli matrices.
One may then for example choose $\theta =0$ (or $\theta =\pi)$.  The
matrix $\CS$ then
becomes block-diagonal and we can ensure ${\cal{H}} = \CS \, {\cal{H}}$ (and, thus,  \eqref{sympinv})
by taking non-zero $Q_0,Q_1$ and $h_1$ (or $h_0$), or non-zero
$P^0,P^1$ and  $h^1$ (or $h^0$).
Then the solution with ${\cal Q}\rightarrow {\cal Q}' = \CS \, {\cal Q}$ is non-supersymmetric.

It is also possible to find $\CR$-matrices mixing the fluxes. Here we content ourselves with giving a field dependent 
example of a matrix $\CR$ fulfilling $\CR^T {\cal{L}} \CR = {\cal{L}}$. For this
one needs ${\cal{L}}$ in \eqref{Lmatrix}
which, together with \eqref{eq:einstein-norm},
can be computed using that the matrices $T$ and $F$ are given by
\begin{equation}
T =  -\frac{1}{2} \left[ \begin {array}{cc} \frac{1}{\Re z} &  1 +
\frac{\bar z}{\Re z} \\
1 +\frac{z}{\Re z} & \frac{|z|^2}{\Re z}  \end {array} \right]  \spa
F = \left[ \begin {array}{cc} 0&-i\\ \noalign{\medskip}-i&0\end {array}
\right] \;,
\end{equation}
with $z=X^1/X^0$.
As an example, we take $\CR$ to be block diagonal in electric/magnetic
fluxes. We find
the solution
\begin{equation}
\CR =  \left( \begin{matrix} {\cal{A}}  & 0  \\  0   & \tilde
{\cal{A}}   \end{matrix} \right) \spa
{\cal{A}} =  \frac{i}{\sqrt{3}} \left( \begin{matrix}    -1
-\frac{\bar z}{\Re z} &   -\frac{1}{\Re z} \\
\frac{|z|^2}{\Re z}    & 1 +   \frac{z}{\Re z}  \end{matrix} \right)
\spa \tilde {\cal{A}}_{IJ} = | \epsilon_{IK} \epsilon_{JL}|
{\cal{A}}_{KL} \;,
\end{equation}
where $\epsilon_{IK}$ is the 2-dimensional Levi-Civita-symbol. 
For this type of  $\CR$-matrix we can then satisfy the symplectic
constraint invariance by either
taking the electric charges $Q_I$ non-zero accompanied by non-zero
fluxes $h_I$ or
magnetic charges $P^I$ together with fluxes $h^I$. 
There are many further possible
$\CR$-matrices that can be constructed, mixing electric and magnetic
charges, even
containing arbitrary parameters.

We also comment on possible $\CS$- and $\CR$-matrices in the STU-model.
A constant (symplectic) $\CS$-matrix satisfying \eqref{SMcom} was
constructed in \cite{Ceresole:2007wx},
\begin{equation}
\CS= {\rm diag}( \epsilon_0, \epsilon_1, \epsilon_2, \epsilon_3,
\epsilon_0, \epsilon_1, \epsilon_2, \epsilon_3) \spa
\epsilon_I = \pm 1
\end{equation}
Without any details of the computation, we also note that it is
possible to find corresponding $\CR$-matrices
mixing the fluxes. A simple example in the case of purely imaginary
$S= X^1/X^0$, $T=X^2/X^0$ and $U=X^3/X^0$ is
\begin{equation}
\CR =  \left( \begin{matrix} {\cal{A}}  & 0  \\  0   & {\cal{A}}^T
\end{matrix} \right) \spa
{\cal{A}} =   \left( \begin{matrix}  \mathbb{I}_2  &  0   \\
0   & {\cal{B}} \end{matrix} \right)  \spa {\cal{B}} = \left(
\begin{matrix}   0 &  \frac{T_2}{U_2}  \\
\frac{U_2}{T_2}   & 0  \end{matrix} \right) \;,
\end{equation}
where $T_2 = \Im T$ and $U_2 = \Im U$.\footnote{We follow the convention to call one of the three scalars $U$, despite the fact that the symbol $U$ also appears in our ansatz for the line element. However, we think that the meaning of $U$ in the following is always obvious from the context.}
It is not difficult to find charge/flux configurations that have the
property that $\CR$ acts non-trivially
on the fluxes, while at the same time leaving the symplectic
constraint invariant.

%%%%%%%%%%%%%%%%%%%%%%%%%%%%%%%%

\subsection{$AdS_2 \times \mathbb{R}^2$ backgrounds  \label{AdS2xR2}}

In the following, we will consider space-times of the type $AdS_2 \times \mathbb{R}^2$,
i.e. line elements \eqref{eq:line-blackb} with constant $A$. In view of the relation \eqref{eq:A-Y},
we thus demand that the $Y^I$ are constant in this geometry. Observe that the latter differs from  the
case of black holes in ungauged supergravity. There, the appropriate $Y^I$ variable is not given in terms of
\eqref{eq:A-psi-U}, but rather in terms of $A = - U$, and the associated flow equations are solved in terms of 
$Y^I = Y^I(r)$ rather than in terms of constant $Y^I$.

For constant $Y^I$, their flow equation yields $q_I = 0$, which implies ${\hat Q}_I = i \, g \,{\rm e}^{2A} 
\, {\hat h}_I$.  Inserting this into the flow equation for $U$ gives
\begin{equation}
{\rm e}^A ({\rm e}^U)' = - 2 g \Im \left[{\rm e}^{i\gamma} \, Y^I \, {\hat h}_I \right] \;,
\label{eq:flow-U-attrac} 
\end{equation}
which equals $({\rm e}^{\psi})'$.  Hence we obtain $\psi' = U'$, 
which is consistent with $A = \psi - U = {\rm constant}$.
Combining the flow equations \eqref{eq:flow-g-W} and \eqref{eq:flow-U-attrac} gives
\begin{equation}
{\rm e}^A \, \left({\rm e}^{U - i \gamma} \right)' = 2 \,i \,  g \, Y^I \, {\hat h}_I = {\rm constant} \;,
\end{equation}
which yields
\begin{equation}
{\rm e}^A \, {\rm e}^{U - i \gamma} = 2 \,i \,  g \, Y^I \, {\hat h}_I \, r + c \;,\;\; \;\;\; c \in \mathbb{C} \;.
\end{equation}
It follows that
\begin{eqnarray}
{\rm e}^A \, {\rm e}^U 
&=& \Re \left[ 2 \,i \,  g \, {\rm e}^{i \gamma}\,
Y^I \, {\hat h}_I \, r + c \, {\rm e}^{i \gamma} \right] \nonumber\\
&=& - 2\,  g \Im \left[ {\rm e}^{i \gamma}\,
Y^I \, {\hat h}_I \right]  r + \Re \left[ c \, {\rm e}^{i\gamma} \right] \;.
\label{eq:valueA-U}
\end{eqnarray}
Now recall that $\gamma$ is given by \eqref{eq:gamma-Z-W}, which takes a constant value, since the $Y^I$
are constant.  Hence $\Re \left[ c \, {\rm e}^{i \gamma} \right] $ is constant,
and it can be removed by a redefinition of $r$, resulting in 
\begin{equation}
{\rm e}^{A+ U} =  - 2\,  g \Im \left[ {\rm e}^{i \gamma}\,
Y^I \, {\hat h}_I \right]  r \;.
\label{eq:A+Uconst}
\end{equation}
Observe that $\Im \Big[{\rm e}^{i \gamma}\,
Y^I \, {\hat h}_I\Big] \neq 0$ to ensure that the space-time geometry contains an $AdS_2$ factor.

Contracting ${\hat Q}_I = i \, g \,{\rm e}^{2A} \, {\hat h}_I$ with $Y^I$ yields the value for ${\rm e}^{2A}$
as
\begin{equation}
{\rm e}^{2A} = - i \, \frac{Y^I \, {\hat Q}_I }{g \,Y^J \,  {\hat h}_J } = 
 - i \, \frac{Z(Y) }{g \, W(Y) } = 
 i \, \frac{{\bar Z}(\bar{Y}) }{g \, \bar W(\bar Y) }
 \;,
 \label{eq:valueA-Z-W}
\end{equation}
and hence
\begin{equation}
 {\hat Q}_I =  \frac{Z(Y) }{ W(Y) } \; {\hat h}_I \;.
\label{eq:attrac-ads2}
\end{equation}
The values of the $Y^I$ are, in principle, obtained by solving \eqref{eq:attrac-ads2}, or equivalently,
\begin{eqnarray}
Q_I - \ft12 (F_{IJ} + \bar{F}_{IJ} ) P^J  &=& \ft12 g \, {\rm e}^{2A} \, N_{IJ} h^J \;, \nonumber\\
- \ft12 N_{IJ} \, P^J &=& g \, {\rm e}^{2A} \, \left(h_I - \ft12 (F_{IJ} + \bar{F}_{IJ} ) h^J \right) \;.
\end{eqnarray}
There may, however, be
flat directions in which case some of the $Y^I$ remain unspecified, and an example thereof is given in section
\ref{sec:exact-sol}. 
The reality of ${\rm e}^{2A}$ forces the phases of $Z(Y)$ and of $W(Y)$ to differ by $\pi/2$ \cite{Dall'Agata:2010gj}.

This relation \eqref{eq:attrac-ads2} has an immediate consequence, similar to the one for black holes
derived in \cite{Dall'Agata:2010gj}.  Namely,
using \eqref{eq:attrac-ads2} and \eqref{eq:valueA-Z-W} in \eqref{eq:symplsimpl} leads to 
\begin{equation}
0 = {\rm Im} \left( \frac{Z(Y) }{ W(Y) } \; {\hat h}_I N^{IJ}  \bar{\hat{h}}_J \right) = {\rm Im} \left( i \, g 
\, {\rm e}^{2A} \; {\hat h}_I N^{IJ}  \bar{\hat{h}}_J \right) = g \, {\rm e}^{2A} \; {\hat h}_I N^{IJ}  \bar{\hat{h}}_J  \ .
\end{equation}
Given that $g \, {\rm e}^{2A} \neq 0$, we infer that for any $AdS_2 \times \mathbb{R}^2$ geometry 
\begin{equation}
{\hat h}_I N^{IJ}  \bar{\hat{h}}_J = 0\ .
\label{eq:attractorconstraint}
\end{equation}
We will use this fact below in section \ref{sec:interpol}. 

%%%%%%%%%%%%%%%%%%%%%%%%%%%%%%%%%%%%%%%%%%%%%%%%%%%%%%

\section{Exact solutions \label{sec:exact-sol}}

In the following, we consider exact (dyonic) solutions of the flow equations in specific
models.  In general, solutions fall into two classes, namely solutions with constant $\gamma$
and solutions with non-constant $\gamma$ along the flow.

\subsection{Solutions with constant $\gamma$}

Let us discuss dyonic solutions in the presence of fluxes. 
For concreteness, we choose $\gamma=0$ in the following.
The flow equations \eqref{eq:flow-U-psi-all} for $U, \psi$ and $A$ read,
\begin{eqnarray} \label{eq:eUpsiA}
({\rm e}^U)' &=& {\rm e}^{- 3 A} \, 
 \Re \left[Y^I \, \hat{Q}_I \right] - g \, {\rm e}^{-A} \, \Im \left[Y^I \, \hat{h}_I \right]
  \;, \nonumber\\
({\rm e}^\psi)' &=& - 2 \, g \, \Im \left[Y^I \, \hat{h}_I \right]\;, \nonumber\\
({\rm e}^A)' &=& - \Re \left[Y^I \, q_I \right] \;, 
\end{eqnarray}
while
the flow equations for the $Y^I$ are
\begin{eqnarray}
\begin{pmatrix} (Y^I - {\bar Y}^I)' \\ (F_I - {\bar F}_I)' 
\end{pmatrix} = 
2 i  \, {\rm e}^{-\psi } \, \left[ - \Im
\begin{pmatrix} 
N^{IK} \, {\hat Q}_K \\ {\bar F}_{IK} \, N^{KJ} \, {\hat Q}_J 
\end{pmatrix}
 + g \, {\rm e}^{2 A} \, \Re
\begin{pmatrix}
N^{IK} \, {\hat h}_K \\ {\bar F}_{IK} \, N^{KJ} \, {\hat h}_J 
\end{pmatrix} \right].
\label{eq:attrac-Q-h-g0}
\end{eqnarray}
In the presence of both fluxes and charges, the flow equations \eqref{eq:attrac-Q-h-g0} cannot be easily
integrated.  A simplification occurs whenever $\Re F_{IJ} = 0$.  This is, for instance, the case
in the model $F = -i X^0 \, X^1$, to which we now turn.

%%%%%%%%%%%%%%%%%%%%%%%%%%%%%%%%%%%

\subsubsection{The $F(X) = - i  X^0 \, X^1$ model}

For this choice of prepotential, we have $F(X) = (X^0)^2 \, {\cal F}(z)$, where ${\cal F}(z) = - i z$ with
$z = X^1/X^0=Y^1/Y^0$. The associated K\"ahler potential reads $K = - \ln (z + \bar z)$, and $z$ is related to 
the dilaton through ${\rm Re} \, z = {\rm e}^{-2 \phi}$. We also have
\begin{eqnarray} \label{Nij}
N_{IJ} = - 2 \begin{pmatrix} 0 & 1 \\ 1 & 0
\end{pmatrix} \;\;\;,\;\;\; N^{IJ} = - \tfrac12 \begin{pmatrix} 0 & 1 \\ 1 & 0
\end{pmatrix} \;,
\end{eqnarray}
as well as 
\begin{eqnarray}
{\cal N}_{00} = - i z \;\;\;,\;\;\; {\cal N}_{11} = - i/z \;\;\;,\;\;\; {\cal N}_{01} = 0 \;.
\end{eqnarray}
The flow equations \eqref{eq:attrac-Q-h-g0} become
\begin{eqnarray}
\begin{pmatrix} (Y^0 - {\bar Y}^0)' \\
(Y^1 - {\bar Y}^1)' \\ 
-i (Y^1 + {\bar Y}^1)' \\
-i (Y^0 + {\bar Y}^0)'
\end{pmatrix} &=& 
 i  \, {\rm e}^{-\psi } \, \Im
\begin{pmatrix} 
\, {\hat Q}_1  \\
{\hat Q}_0
\\  i \, {\hat Q}_0 \\
 i \,  {\hat Q}_1
\end{pmatrix}
 -  i \, g \, {\rm e}^{-\psi + 2 A} \, \Re
\begin{pmatrix}
{\hat h}_1 \\ 
{\hat h}_0 \\
i \,  {\hat h}_0 \\
i \,  {\hat h}_1 
\end{pmatrix} \;.
\label{eq:attrac-Q-h-alpha-01}
\end{eqnarray}
This yields
\begin{eqnarray}
\begin{pmatrix} (Y^0 - {\bar Y}^0)' \\
(Y^1 - {\bar Y}^1)' \\ 
-i (Y^1 + {\bar Y}^1)' \\
-i (Y^0 + {\bar Y}^0)'
\end{pmatrix} &=& 
i  \, {\rm e}^{-\psi } \, \Re 
\begin{pmatrix}
P^0 - g \, {\rm e}^{2A} \, h_1 \\ 
P^1 - g \, {\rm e}^{2A} \, h_0 \\
Q_0 + g\, {\rm e}^{2A} \, h^1 \\
Q_1 + g\, {\rm e}^{2A} \, h^0 \\
\end{pmatrix} \;. 
\label{eq:attrac-alpha-01}
\end{eqnarray}

%%%%%%%%%%%%%%%%%%%%%%%%%%%%%%%%%%%

In order to gain some intuition for finding a solution with all charges and fluxes 
turned on, let us first consider a simpler example. 
We retain only one of the four charge/flux combinations appearing in 
\eqref{eq:attrac-alpha-01}, namely the one with $Q_1 \neq 0, h^0 \neq 0$.
Observe that any of these four combinations satisfies the
symplectic constraint 
\eqref{eq:sympl-constr}.  The associated flow equations \eqref{eq:attrac-alpha-01}
are
\begin{eqnarray}
\begin{pmatrix} (Y^0 - {\bar Y}^0)' \\
(Y^1 - {\bar Y}^1)' \\ 
-i (Y^1 + {\bar Y}^1)' \\
-i (Y^0 + {\bar Y}^0)'
\end{pmatrix} &=& 
i  \, {\rm e}^{-\psi } \,
\begin{pmatrix}
0 \\ 0 \\ 0 \\
Q_1 + g\, {\rm e}^{2A} \, h^0
\end{pmatrix} \;, 
\label{eq:attrac-Q1-h0}
\end{eqnarray}
which yields $Y^1 = {\rm constant}$ and 
\begin{equation}
(Y^0)' = 
- \tfrac12 \, {\rm e}^{-\psi } \, 
\left( Q_1 + g\, {\rm e}^{2A} \, h^0 \right) \;.
\end{equation}
The flow equations for $\psi$ and $A$ read,
\begin{eqnarray}
({\rm e}^\psi)' &=& - 2 g \, 
\Re \left( Y^1 \right) h^0  \; ,
\nonumber\\
\left( {\rm e}^{2 A} \right)' &=& -2 {\rm e}^{-\psi} \Re \left( Y^1 \right) 
 \left(Q_1 + g \, {\rm e}^{2 A} \, h^0  \right) \;.
\end{eqnarray}
For the equation for $A$ we used $q_0=0$ and $q_1={\rm e}^{-A-\psi}\left(Q_1 + g \, {\rm e}^{2 A} \, h^0  \right) $.
The flow equation for $\psi$ can be readily integrated and yields
\begin{equation}
{\rm e}^{\psi} = - 2 g  \,  
\Re \left( Y^1 \right) \, h^0  \, (r + c) \;,
\end{equation}
where $c$ denotes an integration constant. Inserting this into the flow equation for $A$ gives
\begin{equation}
\left( {\rm e}^{2 A} \right)' = 
 \frac{Q_1 + g \, {\rm e}^{2 A} \, h^0}{g \, h^0 \, (r + c)} \;,
\end{equation}
which can be integrated to 
\begin{equation}
{\rm e}^{2 A} = \frac{{\rm e}^{\beta} ( r + c ) - Q_1 }{g \, h^0} \;,
\end{equation}
where $\beta$ denotes another integration constant.
Plugging this into the flow equation for $Y^0$, we can easily integrate the latter,
\begin{equation}
Y^0 = \frac{{\rm e}^{\beta}}{4 g h^0 \, (\Re Y^1)} \, \left(r + \delta \right) \;,
\end{equation}
where $\delta$ denotes a third integration constant. Using \eqref{eq:A-Y}, we infer
\begin{equation}
\delta = c - {\rm e}^{- \beta} Q_1 \;.
\end{equation}
Moreover, for $U$ we obtain
\begin{equation}
{\rm e}^{2 U} = {\rm e}^{2 \psi - 2 A} = \frac{4 g^3 \, (h^0)^3 \, \left(\Re \, Y^1 \right)^2 \, (r + c)^2}{{\rm e}^{\beta} ( r + c ) - Q_1 } \;.
\end{equation}

We take the horizon to be at $r + c = 0$, i.e.\ we set $c =0$ in the following.
Summarizing, we thus obtain
\begin{eqnarray}\label{Q_1h^0sol}
Y^1 &=& {\rm constant} \;, \nonumber\\
Y^0 & =& \frac{{\rm e}^{\beta} \, r - Q_1 }{ 4  \, g h^0 \, (\Re Y^1)}  \;, \nonumber\\
\Re z &=&  \frac{4 \, g \, h^0 \,  (\Re Y^1)^2 }{{\rm e}^{\beta} \, r - Q_1 } \;, 
\nonumber\\
{\rm e}^{2 A} &=& \frac{{\rm e}^{\beta}\, r- Q_1 }{g \, h^0} = 
\frac{4 \, (\Re Y^1)^2 }{\Re z}\;, 
\nonumber\\
{\rm e}^{2 U} &=&\frac{4 g^3 \, (h^0)^3 \, \left(\Re \, Y^1 \right)^2 \, r^2}{{\rm e}^{\beta} \, r - Q_1 } \;, 
\nonumber\\
{\rm e}^{2 \psi} &=& 4 g^2  \,  
\left(\Re \, Y^1 \right)^2 \, (h^0)^2  \, r^2\; .
\end{eqnarray}
We require $h^0>0$ and $Q_1 < 0$ to ensure positivity of $\Re z, {\rm e}^{2 A}$ and ${\rm e}^{2 U}$. This choice is also necessary in order to avoid a singularity of $\Re z$ and ${\rm e}^{2 U}$ at $r ={\rm e}^{-\beta} Q_1$.
The resulting brane solution has non-vanishing entropy density, i.e. ${\rm e}^{2 A(r=0)} \neq 0$.
The imaginary part of $z$ is left unspecified by the flow equations. However, demanding the constraint \eqref{eq:q-Y-real} imposes $Y^1$ to be
real, so that $\Im z = 0$.  Note that $\Re \, Y^1$ corresponds to a flat direction.
The above describes the extremal limit of the solution discussed in sec.\ 7 of 
\cite{Charmousis:2010zz} in the context of AdS/CMT (see also \cite{Taylor:2008tg,Charmousis:2009xr,Gouteraux:2011ce}).\footnote{In order to compare the solutions, one would have to set $\gamma=\delta =1$ in (7.1) of \cite{Charmousis:2010zz}, take their solution to the extremal limit $m^2=\frac{q^2}{8}$ and relate the radial coordinates according to $r^{({\rm us})}=\frac{1}{2l} (r^{({\rm them})})^2 - \frac12 \frac{q}{\sqrt{8}}$.}

%%%%%%%%%%%%%%%%%%%%%%%%%%%%%%%

We now want to solve the equations \eqref{eq:attrac-alpha-01} when all charges and fluxes are non-zero. 
Since the equations are quite difficult to solve directly, we will make an ansatz for ${\rm e}^\psi$ and ${\rm e}^{2A}$ and then solve for the $Y^I$. In the example above we saw 
that ${\rm e}^\psi$ and ${\rm e}^{2A}$ were linear functions of $r$, so we choose the following ansatz for these functions,
\begin{eqnarray}\label{ansatz_psi}
{\rm e}^{\psi(r)}&=& a \, r\,,\nonumber\\
{\rm e}^{2 A(r)}&=& b\, r+c\,.
\end{eqnarray}
When plugging this ansatz into the flow equations \eqref{eq:attrac-alpha-01} we get
\begin{eqnarray}
(Y^0(r))'&=& - \frac{1}{2ar} \left(\bar{\hat{Q}}_1 + i g c \bar{\hat{h}}_1\right)-\frac{g b}{2a} i \bar{\hat{h}}_1\,,\nonumber\\
(Y^1(r))'&=& - \frac{1}{2ar} \left(\bar{\hat{Q}}_0 + i g c \bar{\hat{h}}_0\right)-\frac{g b}{2a} i \bar{\hat{h}}_0\,.
\end{eqnarray}
These equations can easily be integrated to give
\begin{eqnarray} \label{Ysols}
Y^0(r)&=&- \frac{1}{2a} \left(\bar{\hat{Q}}_1 + i g c \bar{\hat{h}}_1\right)\ln r-\frac{g b}{2a} i \bar{\hat{h}}_1r+C^0\,,\nonumber\\
Y^1(r)&=&- \frac{1}{2a} \left(\bar{\hat{Q}}_0 + i g c \bar{\hat{h}}_0\right)\ln r-\frac{g b}{2a} i \bar{\hat{h}}_0r+C^1\,.
\end{eqnarray}
In order for \eqref{ansatz_psi} and \eqref{Ysols} to constitute a solution, several conditions on the parameters $a,b,c$, the charges and the fluxes have to be fulfilled. On the one hand, one has to impose the constraints 
\begin{equation}
{\rm Im}\left(\hat Q_I Y^I\right) = 0\quad , \quad {\rm Re}\left(\hat h_I Y^I\right) = 0 \label{strongconstraints} \ , 
\end{equation}
following from \eqref{eq:flow-U-psi-all} and \eqref{eq:cond-Z-W-Y} for $\gamma=0$. On the other hand, further constraints on the parameters arise from \eqref{eq:A-Y}, \eqref{eq:symplsimpl} and the equation for $\psi$ in \eqref{eq:eUpsiA}. Note that the equation for $U$ does not give additional information, as $U$ is determined, once the $Y^I$ and $\psi$ are given. 

First, the constraints \eqref{strongconstraints} imply
\begin{eqnarray}
{\rm Re}\left( \hat Q_I N^{IJ}  \bar{\hat{h}}_J \right) & = & 0\; , \nonumber \\
{\rm Im} \left( \hat Q_I C^I \right) & = & 0\; , \nonumber \\
{\rm Re} \left( \hat h_I C^I \right) & = & 0\ . \label{strong}
\end{eqnarray}

Next, let us have a look at the equation for $\psi$. Using our ansatz for ${\rm e}^\psi$, it reads
\begin{equation}\label{8param1}
a= - 2 \, g \, \Im \left(Y^I \, \hat{h}_I \right) \,.
\end{equation}
With the form of the $Y^I$ given in \eqref{Ysols} and demanding that the right hand side of \eqref{8param1} is a constant, we obtain the constraint
\begin{equation} \label{hhzero}
\Re \left( \bar{\hat{h}}_1\hat{h}_0 \right) = 0\ .
\end{equation}
This directly leads to a vanishing of the term linear in $r$. Together with the symplectic constraint \eqref{eq:symplsimpl} it also implies that the logarithmic term vanishes. Moreover, we read off
\begin{equation} \label{ach}
a= - 2 g \Im \left(C^I \hat{h}_I \right) \,.
\end{equation}

Next, we analyze the constraints coming from ${\rm e}^{2A}=-N_{IJ}Y^I\bar{Y}^J$. With our ansatz \eqref{ansatz_psi} and the form of $N_{IJ}$ given in \eqref{Nij}, one obtains
\begin{equation}\label{brplusc}
b\,r + c= 4 \Re \left(Y^0\bar{Y}^1 \right) \,.
\end{equation}
This fixes the constant $c$ to be
\begin{equation}
c=4 \Re (C^0\bar{C}^1)\,.
\end{equation}
Using \eqref{ach} the term linear in $r$ on the right hand side of \eqref{brplusc} is just identically $b \, r$, i.e.\ there is no constraint on $b$ (except that it has to be positive in order to guarantee the positivity of ${\rm e}^{2A}$ for all $r$). All other terms on the right hand side vanish if one demands
\begin{eqnarray}
\hat Q_I N^{IJ} \bar{\hat{Q}}_J & = & 0 \ , \label{qqconstr} \\
{\rm Re} \left[ \left(\hat{Q}_I - i g c \hat{h}_I\right) C^I \right] &=& 0\; , \label{reqcconstr} 
\end{eqnarray}
in addition to \eqref{hhzero} and the symplectic constraint \eqref{eq:symplsimpl}. 

We now show that \eqref{reqcconstr} is fulfilled because the even stronger constraint 
\begin{equation} \label{qhrelation}
\hat{Q}_I - i g c \hat{h}_I = 0
\end{equation}
holds. To do so, we note that \eqref{reqcconstr} together with \eqref{strong} and \eqref{ach} imply
\begin{eqnarray}\label{C0C1}
\hat{Q}_0 C^0+\hat{Q}_1 C^1&=&-\frac{ac}{2}\,,\nonumber\\
\hat{h}_0 C^0+\hat{h}_1 C^1&=&-i\frac{a}{2g}\,.
\end{eqnarray}
To get a solution for $C^0$ and $C^1$ one of the two following conditions has to be valid:

\begin{itemize}
\item[i)]If and only if $\hat{Q}_0\hat{h}_1-\hat{h}_0\hat{Q}_1=\rho \neq0$ there is a unique solution for $C^0$ and $C^1$.
\item[ii)]If the two lines in \eqref{C0C1} are multiples of each other then there exists a whole family of solutions.
\end{itemize}

We will now show that case i) can be ruled out. To do so we combine the determinant condition of case i) with the first constraint in \eqref{strong}, i.e.\ we look at the system of linear equations for $\hat{Q}_0$ and $\hat{Q}_1$
\begin{eqnarray}\label{alphabeta13}
\hat{Q}_0\hat{h}_1-\hat{Q}_1\hat{h}_0&=&\rho\,,\nonumber\\
\hat{Q}_0\bar{\hat{h}}_1+\hat{Q}_1\bar{\hat{h}}_0&=&0\,.
\end{eqnarray}
Obviously, the two lines cannot be multiples of each other for $\rho \neq 0$. Thus, in order to find a solution at all, the determinant $\hat{h}_1\bar{\hat{h}}_0+\bar{\hat{h}}_1\hat{h}_0$ has to be non-vanishing. This is, however, in conflict with \eqref{hhzero}. Thus, the two lines in \eqref{C0C1} have to be multiples of each other, implying \eqref{qhrelation}. Note that this implies the absence of the logarithmic terms in the solutions for $Y^I$. 

To summarize we have found the following solution:
\begin{eqnarray}
Y^0(r)&=&-\frac{g b}{2a} i \bar{\hat{h}}_1r+C^0\,,\nonumber\\
Y^1(r)&=&-\frac{g b}{2a} i \bar{\hat{h}}_0r+C^1\, , \nonumber\\
{\rm e}^{\psi(r)}&=& a r\,,\nonumber\\
{\rm e}^{2 A(r)}&=& br+c\,.
\label{eq:exact-sol-gen}
\end{eqnarray}
In addition, the parameters have to fulfill the conditions
\begin{equation}
{\rm Re}(\hat{h}_0\bar{\hat{h}}_1)=0\,,\quad\hat{Q}_I= i g c \hat{h}_I\,,\quad \hat{h}_IC^I=-i\frac{a}{2g}
\,,\quad {\rm Re}(C^0\bar{C}^1)=\frac{c}{4}\,,
\end{equation}
while the parameter $b$ can be any non-negative number. For $b=0$ this solution falls into the class discussed in section \ref{AdS2xR2}. 

Let us finally also mention that one can show that the $F(X) = - i  X^0 \, X^1$ model does not allow for Nernst brane solutions 
(i.e.\ solutions with vanishing entropy) of the first-order equations. Making an ansatz ${\rm e}^U \sim r^\alpha, {\rm e}^{\psi} \sim r^\beta$ and ${\rm e}^{A} \sim r^{\beta\,-\,\alpha}$ for the near-horizon geometry, one can show that the only solutions with both non-vanishing charges 
and fluxes have $\alpha = \beta = 1$ and are captured by the solution discussed in this section after setting $b=0$.  

%%%%%%%%%%%%%%%%%%%%%%%%%%%%%%%

\subsubsection{The $F\,= -\, (X^1)^3/X^0$ model: Interpolating solution between $AdS_4$ and $AdS_2 \times \mathbb{R}^2$ \label{sec:interpol}}

Next, we would like to construct a (supersymmetric)
 solution that interpolates between an $AdS_4$ vacuum at spatial infinity, and
an $AdS_2 \times \mathbb{R}^2$ background with constant $Y^I$ at $r=0$. Thus, asymptotically, we require the solution
to be of the domain wall type (see \eqref{eq:attr-dw}) with $\gamma = \pi/2$. Hence the $Y^I$ satisfy
\eqref{eq:attr-dw-int} with $\alpha^I = \beta_I =0$, so that we may write $Y^I = y^I_{\infty} \, r $ with 
$y^I_{\infty} = g \, N^{IJ} \, \bar{\hat{h}}_J = {\rm constant}$. The latter can be established as follows.
Using $(Y- \bar Y)^I = i g \, h^I \, r $ and $F_I - {\bar F}_I = i g\,  h_I \, r$, we obtain
\begin{equation}
i N_{IJ} \, Y^J = F_I - {\bar F}_{IJ} \, Y^J = i g \, \bar{\hat{h}}_I \, r\;,
\end{equation}
so that
\begin{equation}
\ft12 (Y + \bar Y)^I = g \left( N^{IJ} \, \bar{\hat{h}}_J - \ft12 i \, h^I  \right) r \;.
\end{equation}
It follows that asymptotically,
\begin{equation}
Y^I = g \, N^{IJ} \, \bar{\hat{h}}_J \,  r \,,
\end{equation}
and hence, in an $AdS_4$ background, 
\begin{equation}
{\rm e}^{2A}= {\rm e}^{2U} = - g^2 \, {\hat{h}}_I \,  N^{IJ} \,  \bar{\hat{h}}_J  \; r^2 \;.
\end{equation}
Since this expression only depends on the $Y^I$ through the combinations $N_{IJ}$ and $F_{IJ}$, which 
are homogeneous of degree zero, the $r$-dependence scales out of these quantities, and thus $ {\hat{h}}_I \,  N^{IJ} \,  \bar{\hat{h}}_J$ is a constant. For ${\rm e}^{2A}$ to be positive, we need to require  $ a^2 \equiv  - {\hat{h}}_I \,  N^{IJ} \,  \bar{\hat{h}}_J > 0$.

At $r=0$, on the other hand, we demand $\gamma =\gamma_0 $ as well as $Y^I ={\rm constant}$, with $A$ given by
\eqref{eq:valueA-Z-W}. Thus, we want to construct a solution with a varying $\gamma(r)$ that interpolates between
the values $\pi/2$ and $\gamma_0$.
From \eqref{eq:attractorconstraint} we know that ${\hat{h}}_I \,  N^{IJ} \,  \bar{\hat{h}}_J = 0$ at the horizon.
The $Y^I$ appearing in this expression are evaluated at the horizon.  In general, their values
will differ from the asymptotic values, so that the flow will interpolate
between
an asymptotic $AdS_4$ background satisfying  ${\hat{h}}_I \,  N^{IJ} \,  \bar{\hat{h}}_J < 0$ and an $AdS_2 \times
\mathbb{R}^2$
background satisfying ${\hat{h}}_I \,  N^{IJ} \,  \bar{\hat{h}}_J = 0$. This, however, will not be possible
whenever $F_{IJ}$ is independent of the $Y^I$, such as in the $F = -i X^0 X^1$ model, as already observed in
\cite{Dall'Agata:2010gj}. Thus, in the example below, we will consider the $F\,= -\, (X^1)^3/X^0$ model instead.

The interpolating solution has to have the following properties.  At spatial infinity, 
where $\Im W(Y) = 0$, $\gamma$ is driven away
from its value $\pi/2$ by the term $\Re Z(Y)$ in the flow equation of $\gamma$,
\begin{equation}
\gamma' = \Big[{\rm e}^{2U - 3 \psi} \, \Re Z(Y)\Big]_{\infty}
 = \Big[ {\rm e}^{- 4 U} \, \Re Z(Y) \Big]_{\infty} = \frac{\Re Z(y_{\infty})}{a^4 \,
r^3} \;,
\end{equation}
and hence
\begin{equation}
\gamma (r) \approx \frac{\pi}{2} -  \frac{\Re Z(y_{\infty})}{ 2 a^4 \, r^2} \;.
\end{equation}
Our example below has $\Re Z(y_{\infty})=0$ though and, thus, the phase $\gamma$ will turn out to be constant. 
Near  $r=0$, on the other hand, the deviation from the horizon values can be determined as follows.
Denoting the deviation by
$\delta Y^I = \beta^I \, r^p$, we obtain $\delta ({\rm e}^{2A}) = c \, r^p$ with 
\begin{equation}
 c = - N_{IJ} (\beta^I \, {\bar Y}^J + Y^I \, {\bar \beta}^J) \;,
 \label{eq:devc}
 \end{equation}
where in this expression the $Y^I$ and $N_{IJ}$ are calculated
at the horizon.  Using this, we compute the deviation of ${\bar q}_I$ from its horizon value
${\bar q}_I = 0$, 
\begin{equation}
\delta {\bar q}_I = {\rm e}^{-A - \psi - \mathrm{i} \gamma_0}
 \left(- {\bar F}_{\bar I \bar J \bar L}
(p^J + \mathrm{i} g h^J ) {\bar \beta}^L + \mathrm{i} g \bar{\hat{h}}_I \, c \right) \, r^p\;,
\end{equation}
where all the quantities that do not involve $\beta^I$ are evaluated on the horizon.  The flow equations
for the $Y^I$ then yield
\begin{equation}
p \, N_{IK} \, \beta^K = \frac{{\rm e}^{-i \gamma_0}}{\Delta} \left(- {\bar F}_{\bar I \bar J \bar L}
(p^J + \mathrm{i} g h^J ) {\bar \beta}^L + \mathrm{i} g \bar{\hat{h}}_I \, c \right) \;,
\label{eq:value-beta}
\end{equation}
where
\begin{equation}
\Delta = - 2 g \Im \left({\rm e}^{i \gamma_0} \, \hat{h}_I \, Y^I \right) \;.
\end{equation}
Contracting \eqref{eq:value-beta} with ${\bar Y}^I$ yields $p=1$.  Inserting this value into 
\eqref{eq:value-beta} yields a set of equations that determines the values of the $\beta^I$.

Next, using the flow equation for $\gamma$, we compute the deviation from the horizon value $\gamma_0$, which 
we denote by $\delta \gamma = \Sigma \, r$.  We obtain
\begin{equation}
\Sigma = - \frac{g}{\Delta} \Re \left( {\rm e}^{i \gamma_0} \, \beta^I \, \hat{h}_I \right) \;.
\end{equation}
The example below will have a vanishing $\Sigma$, consistent with a constant $\gamma$.  
Finally, using the flow equation for $\psi$ we compute the change of $\psi$,
\begin{equation}
\delta \psi = - \frac{g}{\Delta} \Im \left( {\rm e}^{i \gamma_0} \, \beta^I \, \hat{h}_I \right) \; r \;.
\label{eq:devpsi}
\end{equation}

%%%%%%%%%%%%%%%%%%%%%%%%%%%%%

We now turn to a concrete example of an interpolating solution. As already mentioned, this will be done in the context of the 
$F(X)=-\frac{(X^1)^3}{X^0}$ model. To obtain an interpolating solution we first need to specify the form of the solution at both ends. 

Let us first have a look at the near horizon $AdS_2 \times \mathbb{R}^2$ region. According to our discussion in sec.\  \ref{AdS2xR2}, we have to solve \eqref{eq:attractorconstraint} and $\hat{Q}_I=ig {\rm e}^{2A}\hat{h}_I$ under the assumption that the $Y^I$ are constant. It turns out that these constraints can be solved, for instance, by choosing $Q_0,P^1,h_1,h^0\neq 0$ and all other parameters vanishing. Of course, due to the symplectic constraint \eqref{eq:sympl-constr}, the four non-vanishing parameters are not all independent, but have to fulfill
\begin{equation}
P^1=\frac{Q_0 h^0}{h_1}\,.
\end{equation}

For $F(Y)=-\frac{(Y^1)^3}{Y^0}$, and introducing  $z=z_1 + i z_2=\frac{Y^1}{Y^0}$, we have
\begin{equation}
N_{IJ}=\begin{pmatrix}
        -4 \,{\rm Im}\left(z^3\right)&6\,{\rm Im}\left(z^2\right)\\
	 6\, {\rm Im}\left(z^2\right)&-12 \,{\rm Im}\left(z\right)
       \end{pmatrix}\,.
\end{equation}
Using 
\begin{equation}
{\rm e}^{2A}=8 |Y^0|^2 z_2^3\,,
\end{equation}
which follows from \eqref{eq:A-Y}, shows that one can fulfill $\hat{Q}_I=ig {\rm e}^{2A}\hat{h}_I$ and \eqref{eq:attractorconstraint} by fixing 
\begin{eqnarray}\label{condabc}
z_1&=&0\,,\nonumber\\
z_2&=&\sqrt{\frac{(3 + 2 \sqrt{3}) h_1}{3 h^0}}\,,\nonumber\\
|Y^0|&=&\frac{3^{\frac{1}{4}} \sqrt{Q_0}}{2 \sqrt{2} z_2^2 \sqrt{h_1}}\, .
\end{eqnarray}
Interestingly we find that the axion has to vanish and all parameters $Q_0,P^1,h_1,h^0$ have to have the same sign. 

In order to describe the asymptotic $AdS_4$ region, we note that asymptotically the charges $Q_0$ and $P^1$ can be neglected and we can read off the asymptotic form of the solution from \eqref{eq:attr-dw-int} with vanishing integration constants $\alpha^I$ and $\beta_I$. More precisely, we have to allow that the asymptotic form of the interpolating solution differs from \eqref{eq:attr-dw-int} by an overall factor. This is because, apriori, we only know that asymptotically $(\psi - 2U)'=0$, cf.\  \eqref{eq:flow-U-psi-all} for vanishing charges. Without the need to match the asymptotic region to a near horizon $AdS_2 \times \mathbb{R}^2$ region, we could just absorb the constant $\psi - 2U$ in a rescaling of the coordinates $x$ and $y$. However, when we start from the hear horizon solution and integrate out to infinity, we can not expect to end up with this choice of convention. Allowing for a non-vanishing constant $\psi - 2U = C \neq 0$, the flow equations \eqref{eq:attr-dw} for $Y^I$ and, thus, also the solutions \eqref{eq:attr-dw-int}, would obtain an overall factor ${\rm e}^C$. Concretely, we obtain
\begin{eqnarray}
Y^0_{AdS_4}(r)&=&i {\rm e}^C \frac{g}{2} h^0 r\,,\nonumber \\
Y^1_{AdS_4}(r)&=&-{\rm e}^C \frac{g \sqrt{h^0 h_1}}{2\sqrt{3}}r\, ,
\label{eq:Y0Y1AdS4}
\end{eqnarray}
with a constant ${\rm e}^C$ to be determined by numerics below. Note, however, that the dilaton ${\rm e}^{-2\phi}=z_2=\sqrt{\frac{h_1}{3h^0}}$ is independent of this factor and it comes out to be real if $h_1$ and $h^0$ have the same sign, consistently with what we found from the near horizon region. 

We would now like to discuss the interpolating solution, which we obtain by specifying boundary conditions at the horizon and then integrating numerically from the horizon to infinity using Mathematica's NDSolve. Given that in the $AdS_4$ region $Y^0$ is purely imaginary and $Y^1$ is purely real, we choose the same reality properties at the horizon, i.e.\ we fulfill \eqref{condabc} by 
\begin{eqnarray}
Y^0_h&=& i \frac{3^{\frac{5}{4}} h^0 \sqrt{Q_0}}{2 \sqrt{2} (3 + 2 \sqrt{3}) h_1^{\frac{3}{2}}}\,,\nonumber\\
Y^1_h&=&i Y^0_h z_2=-\frac{3^{\frac{3}{4}} \sqrt{ h^0 Q_0}}{2 \sqrt{2 (3 + 2 \sqrt{3})} h_1}\,.
\end{eqnarray}
The subscript $h$ means that these values pertain to the horizon. Now we can determine the phase $\gamma_0$ at the horizon using \eqref{eq:gamma-Z-W}. We get
${\rm e}^{-2 i \gamma_0}=-1$ independently of $h_1$, $h^0$, $Q_0$ and $P^1$. We choose $\gamma_0=\pi/2$ as in the asymptotic $AdS_4$ region and we will see that this leads to a constant value for $\gamma$ throughout.

\begin{figure}[th]
\begin{center}
\includegraphics[scale=0.8]{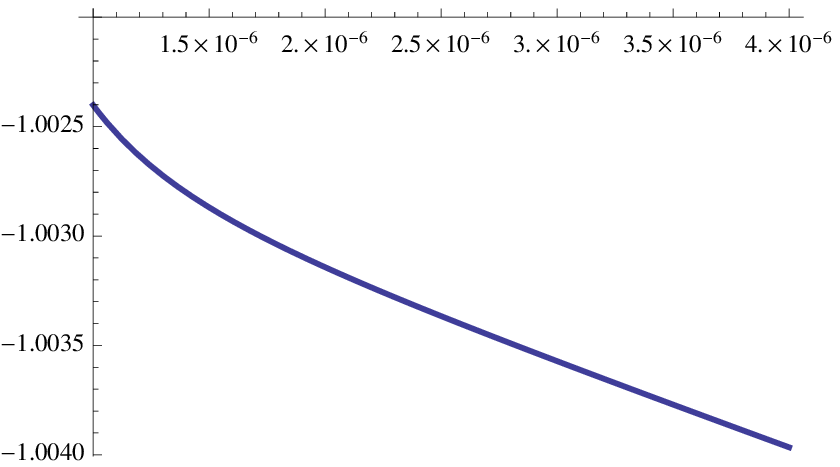} \hskip .7cm
\includegraphics[scale=0.8]{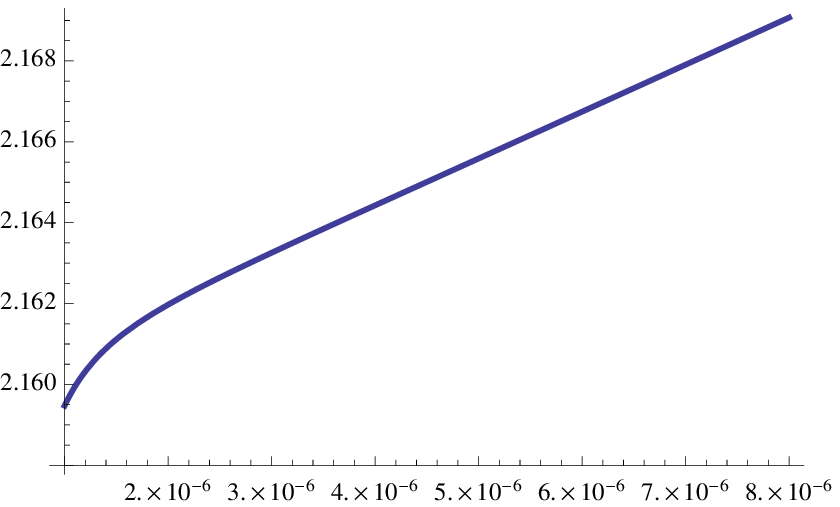}
\begin{picture}(0,0)(0,0)
\put(-405,120){$\Re Y^1(r)$}
\put(-220,105){$r$}
\put(-200,120){$\Im Y^0(r)$}
\put(0,10){$r$}
\end{picture}
\caption{\label{fig1} Interpolating solutions for ${\rm Re}\,Y^1$ and
${\rm Im}\, Y^0$.}
\end{center}
\end{figure}

\begin{figure}[th]
\begin{center}
\includegraphics[scale=0.8]{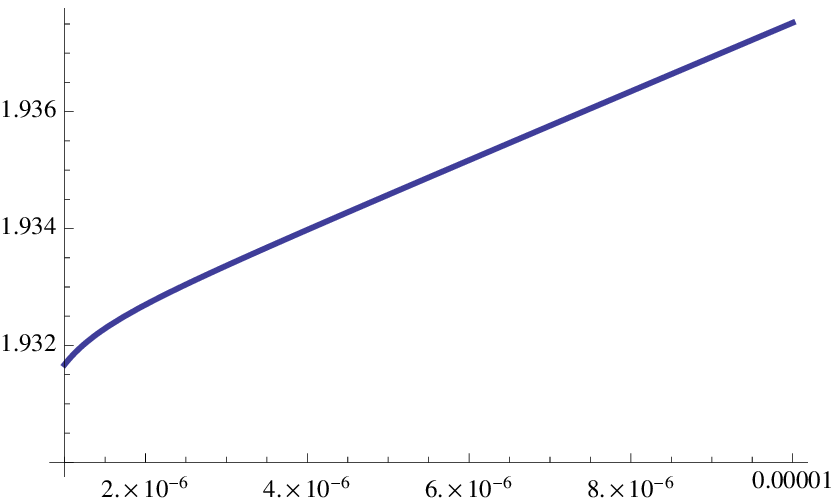} \hskip .7cm
\includegraphics[scale=0.8]{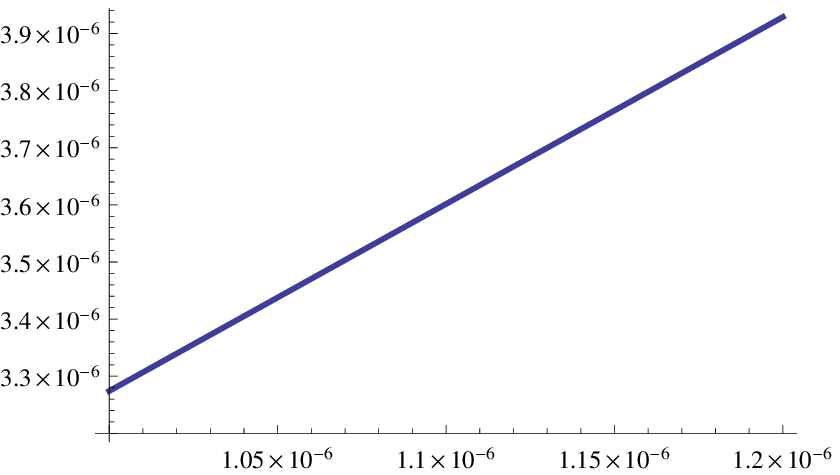}
\begin{picture}(0,0)(0,0)
\put(-405,120){$ {\rm e}^{A(r)} $}
\put(-220,10){$r$}
\put(-180,120){$ {\rm e}^{U(r)} $}
\put(-3,10){$r$}
\end{picture}
\caption{\label{fig2} Interpolating solutions for ${\rm e}^A$ and ${\rm e}^U$.}
\end{center}
\end{figure}

\begin{figure}[th]
\begin{center}
\includegraphics[scale=0.8]{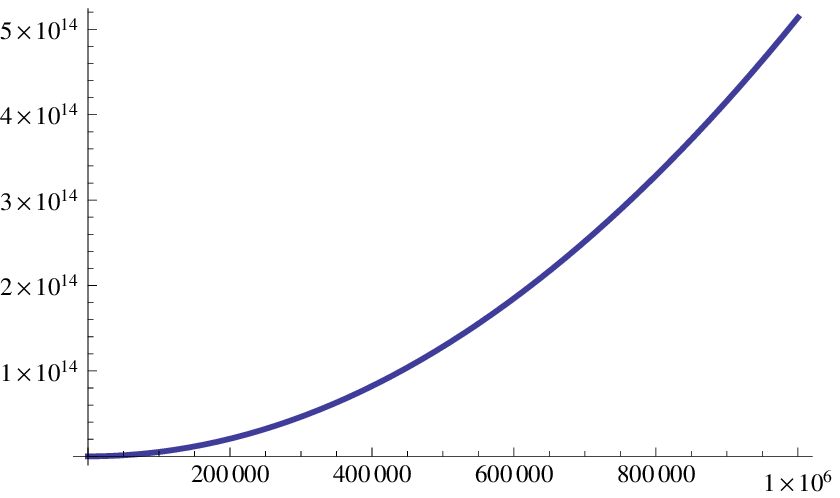} \hskip .7cm
\includegraphics[scale=0.8]{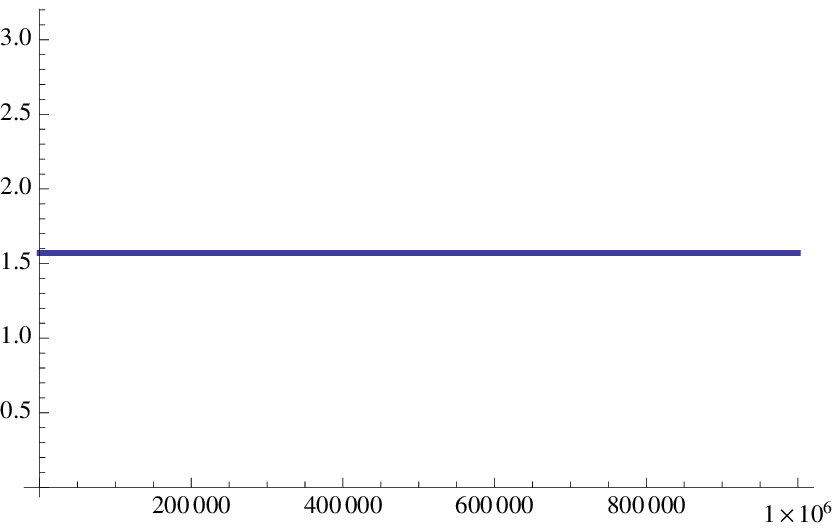}
\begin{picture}(0,0)(0,0)
\put(-400,125){$ {\rm e}^{\psi(r)} $}
\put(-220,10){$r$}
\put(-200,125){$ \gamma (r) $}
\put(-3,10){$r$}
\end{picture}
\caption{\label{fig3} Interpolating solutions for ${\rm e}^\psi$ and $\gamma$.}
\end{center}
\end{figure}

Next we have to determine how the solution deviates near the horizon from $AdS_2 \times \mathbb{R}^2$. We do this following our discussion at the beginning of this section, cf.\ \eqref{eq:devc} - \eqref{eq:devpsi}. Given the reality properties of $Y^I$ in both limiting regions, we choose $\beta^0$ purely imaginary and $\beta^1$ purely real. Solving \eqref{eq:value-beta} for the $\beta$'s results in
\begin{equation}
{\rm Im}\, \beta^0={\rm Re}\,\beta^1\frac{ h^0 \big((\sqrt{3}-2) \sqrt{3 + 2 \sqrt{3}} h_1 \sqrt{\frac{h_1}
    {h^0}} + (5 \sqrt{3}-9) Q_0)\big)}{h_1^2}\,.
\end{equation}
Given this, the deviation of all the fields close to the horizon can be determined (resulting in initial conditions at, say $r_i = 10^{-6}$) and a numerical solution for the flow equations \eqref{eq:flow-U-psi-all}
can be found, integrating from the horizon outwards. To do the numerics we chose 
\begin{equation}
g=1 \quad , \quad {\rm Re}\,\beta^1=0.01 \quad , \quad Q_0 = 1 \quad , \quad  h_1 = 1 \quad , \quad  h^0 = 10\,.
\end{equation}
One can see that asymptotically ${\rm Im}\,Y^0$ and ${\rm Re}\, Y^1$ are linear functions of $r$ (cf. fig. \ref{fig1}), whereas the real part of $Y^0$ and imaginary part of $Y^1$ are zero within the numerical tolerances. The constant ${\rm e}^C$ in \eqref{eq:Y0Y1AdS4} can be determined to be roughly ${\rm e}^C \approx 422$. Futhermore, for $r\to \infty$, the functions ${\rm e}^A$, ${\rm e}^U$ and ${\rm e}^\psi$ behave as expected for the $AdS_4$ background, i.e.\ ${\rm e}^A=\mathcal{O}(r)$, ${\rm e}^U=\mathcal{O}(r)$ and ${\rm e}^\psi=\mathcal{O}(r^2)$ (cf.\ fig.\ \ref{fig2} and \ref{fig3}).\footnote{Note the difference in the plotted $r$-range in fig.\ \ref{fig3} compared to figs.\  \ref{fig1} and \ref{fig2}. Plotting ${\rm Im}\,Y^0$, ${\rm Re}\, Y^1$, ${\rm e}^A$ and ${\rm e}^U$ for larger values would just confirm the linearity in $r$.} Another important feature is that $\gamma(r)=\frac{\pi}{2}$ for any $r$, cf.\ fig.\ \ref{fig3}.

\vskip .5cm

%%%%%%%%%%%%%%%%%%%%%%%%%%%%%%%%%%%%%%%%%%%%%%%

\subsection{Solutions with non-constant $\gamma$}

Next, we turn to solutions with non-constant $\gamma$ along the flow. We focus on two models.  

\subsubsection{The $F\,= -\,i\, X^0\, X^1\,$ model}

We first consider the model $F\,= -\,i\, X^0\, X^1\,$ and 
restrict ourselves to a case involving only the electric charge $Q_0$ and the electric flux $h_0$. 
Since $N^{IJ}$ is off-diagonal, the flow equation for $Y^0$ reads $Y'^0\,=\,0$.  This gives
$Y^0\,=\,C^0$, with $C^0$ a non-vanishing c-number,  which we take to be real, since the phase of $C^0$ can be absorbed
into $\gamma$.  Then, the relation \eqref{eq:A-Y} yields $\exp(2A) = 4 C^0  \, \Re \left({\bar Y}^1 \right)$, from which
we infer that 
$\Re\, Y^1 = \exp (2A) / ( 4 C^0)$.  In the following we take $Q_0, h_0 > 0$ and $C^0 < 0$, for concreteness.

The constraint equation \eqref{eq:cond-Z-W-Y} yields
the following relation between $\gamma$ and $A$,
\begin{equation}
\tan\,[\,\gamma\,]\,=\,\frac{g\,h_0\,}{Q_0}\,{\rm e}^{2A}\,.
\label{eq:tan-g-A-xx}
\end{equation}
Using the $\tau$-variable introduced in \eqref{eq:flow-independent-tau}, 
the flow equation for $A$ can be written as 
\begin{eqnarray}
\dot{A}\,&=&\,-\,\Re\,\Big[\,(\,Q_0 \, {\rm e}^{-\,2A}\,-\,i\,g\,h_0\,)\,C^0\,{\rm e}^{i\gamma}\,\Big] \nonumber\\
&=& \,-\,C^0\,\left(\,Q_0\,{\rm e}^{-2A}\,\cos\,[\,\gamma\,]\,+\,g\,h_0\,\sin\,[\,\gamma\,]\,\right) \;,
\label{eq:flow-A-xx}
\end{eqnarray}
where $\dot{A} = \partial A / \partial \tau$.  Using \eqref{eq:tan-g-A-xx} and (assuming $\gamma \in [-\pi/2, \pi/2]$)
\begin{equation}
\sin\,[\,\gamma\,] = \frac{\tan\,[\,\gamma\,]}{\sqrt{1 + \tan^2[\,\gamma\,]}}
\label{eq:sintan}
\end{equation}
 in \eqref{eq:flow-A-xx} gives
\begin{equation}
\frac{\partial}{\partial \tau} {\rm e}^{2A} =\, - 2 \, C^0 \,\sqrt{Q_0^2\,+\,g^2\,h_0^2\,{\rm e}^{4 A}}\,.
\end{equation}
This is solved by
\begin{equation}
{\rm e}^{2A}\,=\, \alpha  \left( {\rm e}^{ \delta \, \tau} - \frac{Q_0^2}{4\,  \alpha^2 \, g^2 \, h^2_0 } \, 
{\rm e}^{ -\delta \, \tau} \right) \;,
\label{eq:sol-A-tau}
\end{equation}
where $\delta = - 2 \, g \, h_0 \, C^0 > 0$ and 
$\alpha$ denotes an integration constant. Setting $\alpha = Q_0/(2 g h_0)$ this becomes 
\begin{equation}
{\rm e}^{2A}\,=\, \frac{Q_0}{g h_0}\, \sinh(\, \delta \, \tau\,) \;,
\label{eq:Atausinh}
\end{equation}
and demanding
$\exp (2 A)$ to be positive restricts the range of $\tau$ to lie between $0$ and  $\infty$.

The flow equation for $\psi$ is given by 
\begin{equation}
\dot{\psi}=\delta\, \sin \,[\,\gamma\,] = \delta\, \tanh (\,\delta \, \tau\,)\,,
\end{equation}
where we used \eqref{eq:sintan}, \eqref{eq:tan-g-A-xx} and \eqref{eq:Atausinh}. This can easily be solved by
\begin{equation}
{\rm e}^{\psi - \psi_0}\,= \,  \cosh (\,\delta \, \tau\,)\;,
\end{equation}
where $\psi_0$ denotes an integration constant which we set to zero.  Using $dr = \exp ( \psi) \, d \tau$, we
establish
\begin{equation}
r\,-\,r_0= \frac{\sinh(\,\delta \, \tau\,)}{\delta} > 0\;.
\end{equation}
As $\tau \rightarrow 0$ we have $\exp(2A) \rightarrow 0$ and $r - r_0 \rightarrow 0$.  

The dilaton is given by 
\begin{equation}
\Re z = \,\Re\left( \,\frac{Y^1}{Y^0}\,\right) \,=\,\frac{{\rm e}^{2A}}{4\,{(C^0)}^2}\,.
\end{equation}
We notice here that the dilaton $e^\phi = (\sqrt{\Re z})^{-1}$ blows up at $r \rightarrow r_0$. The same happens with the curvature invariants like the Ricci scalar. As the curvature invariants vanish asymptotically for large $r$ (we checked this up to second order in the Riemann tensor), one could still hope that our solution describes the asymptotic region of a global solution, once higher curvature corrections are taken into account. 

%%%%%%%%%%%%%%%%%%%%%%%%%%%%%%%%%%%%

\subsubsection{The STU model \label{sec:STUmodel}}

Now we consider
 the STU-model, which is based on $F(Y) = - Y^1 Y^2 Y^3/Y^0$.  We denote the $z^i = Y^i/Y^0$
(with $i = 1,2,3$) by $z^1 = S \,,\, z^2 = T \,,\, z^3 = U$.
For concreteness, we will only consider solutions that are supported by an electric charge $Q_0$ and an electric flux $h_0$.  In addition, we will restrict ourselves to axion-free 
solutions, that is solutions with vanishing $\Re S \,,\, \Re T \,,\, \Re U$.  The solutions will thus be supported
by $S_2 = \Im S \,,\, T_2 = \Im T \,,\, U_2 = \Im U$, so that
\begin{eqnarray}
N_{IJ} &=& \begin{pmatrix}
 4 S_2 T_2 U_2 & 0 & 0 & 0\\
 0 & 0 & -2 U_2 & - 2 T_2  \\
 0 & -2 U_2 & 0 & - 2 S_2 \\
 0 & - 2 T_2 & - 2 S_2 & 0
 \end{pmatrix}\,, \nonumber \\
 N^{IJ} &=& \frac{1}{4 S_2 T_2 U_2} \begin{pmatrix}
 1 & 0 & 0 & 0\\
 0 & S_2^2 & - S_2 T_2 & - S_2 U_2 \\
 0 & - S_2 T_2 & T_2^2 & - T_2 U_2 \\
 0 & - S_2 U_2 & - T_2 U_2 & U_2^2
 \end{pmatrix}\,.
 \label{eq:matrix-N-stu}
 \end{eqnarray}

The flow equations for
the $Y^i$ imply that they are constant. To ensure that 
$S, T$ and $U$ are axion-free, we take the constant $Y^i$ to be
purely imaginary, i.e.\ $Y^i\,=\,i\,C^i$, where the $C^i$ denote real constants. 
Using \eqref{eq:A-Y} in the form
\begin{equation}
{\rm e}^{2A}\,= 8\,|Y^0|^2\,S_2\,T_2\,U_2 \;,
\label{eq:rel-A-Y-stu}
\end{equation}
the flow equation \eqref{eq:flow-independent-tau} for $Y^0$ gives
\begin{equation}
\dot{Y}^0\,=\,2\,|Y^0|^2\,(\,Q_0\,{\rm e}^{-\,2\,A}\,+\,i\,g\,h_0\,)\,{\rm e}^{-\,i\,\gamma\,}.
\label{eq:dot-Y0-gam}
\end{equation}
Using the constraint equation \eqref{eq:cond-Z-W-Y} in the form
\begin{equation}
\Im\,\Big[\,(Q_0\,{\rm e}^{-\,2 A}\,+\,i\,g\,h_0\,)\,{\rm e}^{-\,i\gamma\,}\,\bar{Y}^0\,\Big]\,=\,0\,
\label{eq:const-q-h-g}
\end{equation}
as well as the
flow equation for $A$, 
\begin{equation}
\dot{A}\,=\,-\,\Re\,\Big[\left(Q_0\,{\rm e}^{-\,2 A}\,+\,i\,g\,h_0\,\right)\,{\rm e}^{-\,i \gamma\,}\,\bar{Y}^0\,\Big]\,,
\label{eq:flow-A-dot}
\end{equation}
we can rewrite \eqref{eq:dot-Y0-gam} as
\begin{equation}
\dot{Y}^0\,=\,-\,2\,Y^0\,\dot{A}\,.
\end{equation}
This immediately gives 
\begin{equation}
Y^0\,=\,C^0\,{\rm e}^{-\,2A}\,,
\end{equation}
where $C^0$ denotes an integration constant which we take to be real, 
for simplicity. Then, it follows from 
\eqref{eq:rel-A-Y-stu} that
\begin{equation}
8\,\frac{C^1\,C^2\,C^3}{C^0}\,=\,1 \;.
\end{equation}
For concreteness we take $C^I < 0$ ($I=0, \dots, 3$) in the following (ensuring the positivity of $S_2, T_2$ and $U_2$), as well as $h_0,Q_0 > 0$.

The constraint equation \eqref{eq:const-q-h-g} yields
the following relation between $\gamma$ and $A$,
\begin{equation}
\tan\,[\,\gamma\,]\,=\,\frac{g\,h_0\,}{Q_0}\,{\rm e}^{2A}\,.
\label{eq:tan-g-A}
\end{equation}
The flow equation \eqref{eq:flow-A-dot}
can be written as 
\begin{equation}
\tfrac12 \, \frac{\partial}{\partial \tau} {\rm e}^{2A} = 
\,-\,C^0\,\left(\,Q_0\,{\rm e}^{-2A}\,\cos\,[\,\gamma\,]\,+\,g\,h_0\,\sin\,[\,\gamma\,]\,\right)
\end{equation}
which, using \eqref{eq:tan-g-A}, leads to 
\begin{equation}
\tfrac14 \, \frac{\partial}{\partial\tau} {\rm e}^{4A} =- C^0 \,\sqrt{Q_0^2\,+\,g^2\,h_0^2\,{\rm e}^{4 A}}\,.
\end{equation}
This can be solved to give 
\begin{equation}
{\rm e}^{4A}\,=\,\frac{4\,g^4\,h_0^4 \,\left(C^0\right)^2\,(\tau\,+\,c\,)^2\,-\,Q_0^2\,}{g^2\,h_0^2} \;,
\label{e4A}
\end{equation}
where $c$ denotes a further integration constant.
Hence, the prefactor of the planar part of the metric is 
\begin{equation}
{\rm e}^{2A}\,=\,\frac{\sqrt{4\,g^4\,h_0^4 \,\left(C^0\right)^2\,(\tau\,+\,c\,)^2\,-\,Q_0^2\,}}{g\,h_0}\,.
\label{eq:e2atau}
\end{equation}
This is well behaved provided that $(\tau + c)^2 \geq Q_0^2/(4 g^4 \, h_0^4 \, (C^0)^2)$.

The flow equation for $\psi$,
\begin{equation}
\dot{\psi}\,=\,-2\,g\,h_0\,C^0\,{\rm e}^{-\,2A}\,\sin\,[\,\gamma\,]\, 
\end{equation}
is solved by (using \eqref{eq:sintan}, \eqref{eq:tan-g-A} and \eqref{e4A})
\begin{equation}
{\rm e}^{\psi}\,=\,\frac{1}{\tau\,+\,c}\, \;.
\end{equation}
The radial coordinate is then related to the $\tau$ variable by 
\begin{equation}
r\,[\,\tau\,]\,=\,\log\,[\,\tau\,+\,c\,]\,,
\end{equation}
where we set an additional integration constant to zero.  

The physical scalars are given by 
\begin{equation}
S_2\,=\,\frac{C^1}{C^0}\,{\rm e}^{2A} \;\;\;,\;\;\; T_2\,=\,\frac{C^2}{C^0}\,{\rm e}^{2A} \;\;\;,\;\;\; U_2\,=\,\frac{C^3}{C^0}\,{\rm e}^{2A} \;.
\end{equation}
They are positive as long as ${\rm e}^{2A}$ is. However, as in the previous example, they vanish at the lower end of the radial coordinate (indicating that string loop and $\alpha'$ corrections should become important). Again also the curvature blows up there and one can at best consider this solution as an asymptotic approximation to a full solution which might exist after including higher derivative terms. As we said in the introduction, it would be worthwhile to further pursue the search for everywhere well behaved solutions with non-constant $\gamma$ as they might be radically different from the ungauged case. 

%%%%%%%%%%%%%%%%%%%%%%%%%%%%%%%%%%%%%%%%%%%%%%%%%%

\section{Nernst brane solutions in the STU model \label{sec:nernst}}

In the following we construct Nernst brane solutions (i.e.\ solutions with vanishing entropy density) in a particular model, namely the STU-model already discussed in sec.\ \ref{sec:STUmodel}.  As there, we denote the $z^i = Y^i/Y^0$
(with $i = 1,2,3$) by $z^1 = S \,,\, z^2 = T \,,\, z^3 = U$.  
For concreteness, we will only consider solutions that are supported by the electric charge $Q_0$ and the electric fluxes $h_1, h_2, h_3$.  In addition, we will restrict ourselves to axion-free  solutions, that is
solutions with vanishing $\Re S \,,\, \Re T \,,\, \Re U$.  The solutions will thus be supported
by $S_2 = \Im S \,,\, T_2 = \Im T \,,\, U_2 = \Im U$, and the corresponding matrices $N_{IJ}$ and $N^{IJ}$ are given in eq.~\eqref{eq:matrix-N-stu}. 
We thus have that  ${\rm e}^{2A}$ is determined by  \eqref{eq:rel-A-Y-stu}. 
In the following, we take $Y^0$ to be real, so that the $Y^i$ will be purely imaginary.

Let us consider the flow equation \eqref{eq:flow-independent-tau}
for the $Y^I$.
We set $\gamma =0$. Instead of working with a  $\tau$ coordinate defined by $d \tau = {\rm e}^{- \psi} \, dr$,
we find it convenient to work with 
$d \tau = - {\rm e}^{- \psi} \, dr$.
We obtain
\begin{eqnarray}
\dot{Y}^0\,&=& - \frac{Q_0}{4 S_2 T_2 U_2} \;, \nonumber\\
\dot{Y}^i\,&=&\, 2\,i\,g\,Y^i\,\Big[\,2\,Y^i\,h_i\,-\,Y^j\,h_j\,\Big]\,,
\end{eqnarray}
where $i,j=1,2,3$ and $\dot{Y}^I = \partial Y^I/\partial \tau$.
Here, $i$ is not being summed over, while $j$ is.
The flow equations for $Y^i$ are solved by
\begin{equation}
Y^i\,= - \frac{i}{ 2 \ g\,h_i \, \tau} \;. 
\label{eq:sol-Yi-nernst}
\end{equation}
Next, using that for an axion-free solution
$S_2\,=\,-\,i\, Y^1/Y^0$, $T_2\,=\,-\,i\,Y^2/Y^0$ and $U_2\,=\,-\,i\,Y^3/Y^0$, 
and inserting \eqref{eq:sol-Yi-nernst} into the flow equation for $Y^0$, we obtain
\begin{equation}
\dot{Y}^0\,=  2 g^3 \, Q_0\, h_1 h_2 h_3 \,\left(Y^0\right)^3 \, \tau^3 \;.
\end{equation}
This equation can be easily solved to give
\begin{equation}
Y^0\,=-\frac{1}{\sqrt{- g^3 \, Q_0\,h_1 \, h_2 \, h_3 \, (\,\tau^4+\,C^0\,)\,}}\,,
\end{equation}
where $C^0$ denotes an integration constant. We also take  $Q_0<0$ and $h_1, h_2, h_3>0$. This ensures that the physical
scalars 
\begin{eqnarray}
S_2 & =&\frac{1}{2\,g\,h_1\,\tau}\,\sqrt{-g^3 \, \,Q_0\, h_1\,h_2\,h_3 \,(\,\tau^4+\,C^0)\,}\,, \nonumber \\
T_2&=& \frac{1}{2\,g\,h_2\,\tau}\,\sqrt{-g^3 \, Q_0 \, h_1\,h_2\,h_3 \,(\,\tau^4+\,C^0)\,}\, ,\nonumber \\
U_2&=&\frac{1}{2\,g\,h_3\,\tau}\,\sqrt{-g^3 \, Q_0 \, h_1\,h_2\,h_3 \,(\,\tau^4+\,C^0)\,}\, 
\end{eqnarray}
take positive values.

Next, we consider the flow equation for $\psi$ following from \eqref{eq:flow-U-psi-all}.  Using
\eqref{eq:sol-Yi-nernst}, we obtain
\begin{equation}
\dot{\psi}\,=\,2\,g\,\Im\,\Big[\,Y^i\,h_i \Big]\,=\,- \frac{3}{\tau}\,,
\end{equation}
which upon integration yields
\begin{equation}
{\rm e}^{\psi -\psi_0}\,=\,\frac{1}{\tau^3} \;,
\end{equation}
where $\psi_0$ denotes an integration constant, which we set to $\psi_0 =0$. We can thus take $\tau$ to range between $0$ and $\infty$.
The relation $d \tau = - {\rm e}^{- \psi} \, dr$ then results in 
\begin{equation}
r\,[\,\tau\,]\,=\,\frac{1}{2 \, \tau^2}\,+\,C_r \;,
\end{equation}
where $C_r$ denotes a real constant, which we take to be zero in the following.
This sets the range of $r$ to vary from $\infty$ to $0$ as 
$\tau$ varies from $0$ to $\infty$.

Using \eqref{eq:rel-A-Y-stu}, we obtain
\begin{equation}
{\rm e}^{2A}\,=
\,\frac{ \sqrt{-Q_0}}{\sqrt{g^3\,h_1\,h_2\,h_3}}\,\frac{\sqrt{\tau^4\,+\,C^0}}{\tau^3}\,,
\end{equation}
and hence
\begin{equation}
{\rm e}^{-2U}\,= 
\,\frac{ \sqrt{-Q_0}}{\sqrt{g^3\,h_1\,h_2\,h_3}} \, \sqrt{\tau^4\,+\,C^0} \; \tau^3 \,.
\end{equation}
We take $C_0$ to be non-vanishing and positive to ensure that the curvature scalar remains finite througout
and that the above describes
a smooth solution of Einstein's equations.

Asymptotically ($\tau \rightarrow 0$) we obtain
\begin{eqnarray}
{\rm e}^{2A} &\stackrel{r\rightarrow \infty}{\longrightarrow}& \alpha \left(1 + \frac{1}{8 \, C^0 \, r^2} \right) \left(2 \, r \right)^{3/2} \;,
\nonumber\\
{\rm e}^{-2U} &\stackrel{r\rightarrow \infty}{\longrightarrow}& \alpha \left(1 + \frac{1}{8 \, C^0 \, r^2} \right) \left(\frac{1}{2 \, r}\right)^{3/2} \;,
\nonumber\\
{\rm e}^{2U} &\stackrel{r\rightarrow \infty}{\longrightarrow}& \alpha^{-1} \left(1 - \frac{1}{8 \, C^0 \, r^2} \right) \left(2 \, r \right)^{3/2} \;,
\end{eqnarray}
where
$\alpha =  \sqrt{-Q_0 \, C^0/(g^3 \, h_1\,h_2\,h_3)}$.  Observe that both $\exp(2A)$ and $\exp(2U)$ have
an asymptotic $r^{3/2}$ fall-off, which is rather unusual. 
Finally, observe that 
the scalar fields $S_2, T_2$ and $U_2$ grow as $\left( C^0 \, r\right)^{1/2}$ asymptotically.

The near-horizon solution (which corresponds to $\tau \rightarrow \infty$) can be obtained by setting $C^0=0$
in the above.  This yields
\begin{eqnarray}
{\rm e}^{2A}\,&\stackrel{r\rightarrow 0}{\longrightarrow}& 
\,\frac{ \sqrt{-Q_0}}{\sqrt{g^3\,h_1\,h_2\,h_3}}\,\left(2 \, r \right)^{1/2}\,, \nonumber\\
{\rm e}^{-2U}\,&\stackrel{r\rightarrow 0}{\longrightarrow}& 
\,\frac{ \sqrt{-Q_0}}{\sqrt{g^3\,h_1\,h_2\,h_3}} \, \frac{1}{(2 \, r )^{5/2}} \,,
\end{eqnarray}
while $S_2, T_2$ and $U_2$ approach the horizon as $r^{-1/2}$. The near-horizon geometry has an infinitely long radial throat as well as a vanishing area
density, which indicates that the
solution describes an extremal black brane with vanishing entropy density.
It describes a supersymmetric Nernst brane solution that is valid in supergravity, and hence in the 
supergravity approximation to string theory when switching off
string theoretic $\alpha'$-corrections.  When embedding
this solution into type IIA string theory, 
where $S, T$ and $U$ correspond to K\"ahler moduli,
the fact that asymptotically and at the horizon $S_2 \,,\, T_2$ and $U_2$ blow up indicates
that this solution should be viewed as a good solution only in ten dimensions.  In the heterotic description, 
where $S_2$ is related to the inverse of the heterotic string coupling $g_s$,  
the behaviour $S_2 = \infty$ (which holds asymptotically and at the horizon) is consistent with working in the classical limit $g_s \rightarrow 0$.\footnote{In fact, by analyzing this solution in the STU model viewed as arising from heterotic compactification on $K3\,\times\, T^2$, one can explicitly verify that (for appropriately chosen flux parameters) the Nernst solution is a smooth and consistent solution of the 10D supergravity action arising as the low-energy limit of the heterotic string. In particular, the 10D string coupling is given by $g^2_{10\,s}\,=\,V_{K3}\,\frac{h_1}{h_3}$ and, thus, can be made small for appropriate choices of $h_1$ and $h_3$, even for values of the $K3$ volume $V_{K3}$ which are large enough to allow the neglect of $\alpha'$-corrections.}

We observe that the Nernst brane solution constructed above differs substantially from the one
constructed in \cite{Goldstein:2009cv}.  This is related to the fact that here we deal with a flux potential
in gauged supergravity, and not with a cosmological constant as in \cite{Goldstein:2009cv}.

The solution given above is supported by electric charges and fluxes, only.
Additional supersymmetric Nernst solutions can
be generated through the technique of symplectic transformations.  On the other hand, a  non-supersymmetric Nernst solution can be obtained
from the above one by replacing the charge vector $\cal Q$ by ${\cal S}\, \cal Q$ in the flow equations, as described in section 
\ref{sec:gen-first-order}.

%%%%%%%%%%%%%%%%%%%%%%%%%%%%%%%%%%%%%%%%%%%%

\section{Non-extremal deformation \label{sec:deform}}

In this section we construct a class of non-extremal black brane solutions that are based on
first-order flow equations.  We do this by turning on a deformation parameter in the line element describing
extremal black branes.  In the context of $N=2 \; U(1)$ gauged supergravity in five dimensions, it was shown
in \cite{Cardoso:2008gm} that there exists a class of non-extremal charged black hole solutions that are based on first-order
flow equations, even though their extremal limit is singular.  There, the non-extremal parameter is not only encoded
in the line element, but also in the definition of the physical charges.  Here, we will only study the case
where the deformation parameter enters in the line element. A different procedure for constructing non-extremal black hole
solutions has recently been given in \cite{Galli:2011fq} in the context of ungauged supergravity in four dimensions.

We begin by deforming the line element \eqref{eq:line-blackb} into
\begin{equation}
ds^2 = - ( {\rm e}^{2 U} - \mu \, {\rm e}^{2 R})  dt^2 +  ( {\rm e}^{2U} -  \mu \, {\rm e}^{2 R})^{-1} dr^2 
+ {\rm e}^{2(\psi-U)}
\left( dx^2 + dy^2 \right) \;,
\label{eq:line-def}
\end{equation}
where $R = R(r)$ and $\mu$ is the non-extremal deformation parameter.
Using
\begin{eqnarray}
- \frac12 \, \sqrt{-g} \,  \mathfrak{R} &=&
{\rm e}^{2 \psi} \left[\left(U' - \psi'\right)^2 + 2 \psi'^2 + 2 \psi'' - U'' \right]
\nonumber\\
&& + \mu \, {\rm e}^{2 (\psi - U + R)} \left[     
6 U' \,\psi' - 3  U'^2 - 3  \psi'^2 + 2 U''  -2 \psi'' \right. \nonumber\\
&& \left. \qquad \qquad \qquad \qquad - 2 R'^2 - 4 R' \, (\psi' - U' ) - R''
\right] \;, \label{Rdeform}
\end{eqnarray}
the one-dimensional action \eqref{eq:action-sympl}
gets deformed into
\begin{eqnarray}
- S_{1d} &=& - S_{1d}(\mu=0) + \mu \, \int dr \, {\rm e}^{2(\psi-  U + R )} \left[ (\psi' -  U')^2 + 2 R' \, 
(\psi' - U' )
\right] \nonumber\\
&& - \mu \, \int dr \, {\rm e}^{2 (R + \psi - U - A) } \,  \, N_{IJ} \left(Y'^I - A' \, Y^I \right) \left( {\bar Y}'^J 
- A' \, {\bar Y}^J \right) \nonumber\\
&&+ \mu \, \int dr \, \frac{d}{dr} \left[{\rm e}^{2(\psi - U + R )} \left(2 U' - 2 \psi' - R' \right) \right] \;,
\end{eqnarray}
where $S_{1d}(\mu=0)$ denotes the action given in \eqref{eq:action-sympl}.  Using \eqref{eq:A-Y}
and \eqref{eq:relAp-Y}, 
the second line of this action can be written as
\begin{eqnarray}
&&- \mu \, \int dr \, {\rm e}^{2 (R + \psi - U - A) } \,  \, \left[ N_{IJ} \, Y'^I \, {\bar Y}'^J + (A')^2 \, {\rm e}^{2A}
\right] \nonumber\\
&& = - \mu \, \int dr \, {\rm e}^{2 (R + \psi - U - A) } \,  \, \Big[ N_{IJ} \, \left(Y'^I - {\rm e}^A \, N^{IK} \, 
{\bar q}_K \right) 
\left({\bar Y}'^J - {\rm e}^A \, N^{JL} \, q_L \right) \\
&& 
\qquad \qquad \qquad  
+ {\rm e}^{A} \left[ q_I \left(Y'^I - \ft12 {\rm e}^A \, N^{IK} \, {\bar q}_K \right)
+ {\bar q}_I \left({\bar Y}'^I - \ft12 {\rm e}^A \, N^{IL} \, q_L \right) \right]
 + (A')^2 \, {\rm e}^{2A}
\Big] \;. \nonumber
\end{eqnarray}
Using this, we rewrite the action (up to a total derivative) as 
\begin{eqnarray}
- S_{1d} &=& - S_{1d}(\mu=0) \nonumber\\
&& + \mu \, \int dr \, 
{\rm e}^{2R} \left( \left[ \left({\rm e}^{\psi - U}\right)' - \left( {\rm e}^A \right)' \right]^2 
+ 2 \left({\rm e}^A \right)' \left[\left({\rm e}^{\psi - U} \right)' - \left({\rm e}^A \right)' \right] \right)
\nonumber\\
&& + \mu \, \int dr \, \ft12 \left( {\rm e}^{2 R} \right)' \left( {\rm e}^{2 (\psi - U) } \right)'
\nonumber\\
&& - \mu \, \int dr \, {\rm e}^{2 (R + \psi - U - A) } \,  \, \Big[ N_{IJ} \, \left(Y'^I - {\rm e}^A \, N^{IK} \, 
{\bar q}_K \right) 
\left({\bar Y}'^J - {\rm e}^A \, N^{JL} \, q_L \right) \nonumber\\
&& 
\qquad \qquad \qquad  
+ {\rm e}^{A} \left[ q_I \left(Y'^I - \ft12 {\rm e}^A \, N^{IK} \, {\bar q}_K \right)
+ {\bar q}_I \left({\bar Y}'^I - \ft12 {\rm e}^A \, N^{IL} \, q_L \right) \right] \Big] \nonumber\\
&& + \mu \, \int dr \, {\rm e}^{2R} \Big[ \left({\rm e}^A \right)' \Big]^2 \left( 1 - {\rm e}^{2 (\psi - U - A) } \right) \;.
\label{eq:action-mu-def}
\end{eqnarray}
Now consider varying the action \eqref{eq:action-mu-def} and imposing the first-order flow equations 
\eqref{eq:flow-U-psi-all} on the fields.  We begin by considering the variation
\begin{eqnarray}
\delta
\left[ q_I \left(Y'^I - \ft12 {\rm e}^A \, N^{IK} \, {\bar q}_K \right) \right]
+ \delta \left[{\bar q}_I \left({\bar Y}'^I - \ft12 {\rm e}^A \, N^{IL} \, q_L \right) \right] \;,
\end{eqnarray}
which on a solution to the first-order flow equations equals
\begin{eqnarray}
q_I \, \delta (Y'^I) + {\bar q}_I \, \delta ({\bar Y}'^I ) - q_I \, N^{IJ} \, {\bar q}_J \, \delta \left({\rm e}^{A}
\right)
- q^I \, \delta ( N^{IJ} ) \, {\bar q}_J \, {\rm e}^A  \;,
\end{eqnarray}
showing that there is no need to explicitly compute the variation of $q_I$. Note, in particular, that
the variation of $\gamma$ drops out of the problem.  The independent variations are
\begin{eqnarray}
&& \delta_{\psi}
\left[ q_I \left(Y'^I - \ft12 {\rm e}^A \, N^{IK} \, {\bar q}_K \right) \right]
+ \delta_{\psi} \left[{\bar q}_I \left({\bar Y}'^I - \ft12 {\rm e}^A \, N^{IL} \, q_L \right) \right] = 0 \;, 
\nonumber\\
&& \delta_{U}
\left[ q_I \left(Y'^I - \ft12 {\rm e}^A \, N^{IK} \, {\bar q}_K \right) \right]
+ \delta_{U} \left[{\bar q}_I \left({\bar Y}'^I - \ft12 {\rm e}^A \, N^{IL} \, q_L \right) \right] = 0 \;, 
\nonumber\\
&& \delta_{Y}
\left[ q_I \left(Y'^I - \ft12 {\rm e}^A \, N^{IK} \, {\bar q}_K \right) \right]
+ \delta_{Y} \left[{\bar q}_I \left({\bar Y}'^I - \ft12 {\rm e}^A \, N^{IL} \, q_L \right) \right] = 
q_I \, \delta (Y'^I) \nonumber\\
&& \qquad \qquad \qquad 
- q_I \, N^{IJ} \, {\bar q}_J \, \delta_Y \left({\rm e}^{A}
\right)
- i \, q^I \, N^{IP} \, F_{PQL} \, N^{QJ} \, {\bar q}_J \, {\rm e}^A \, \delta Y^L \;.
\end{eqnarray}
Then, varying the action \eqref{eq:action-mu-def} with respect to $\psi$ gives
\begin{eqnarray}
\left( {\rm e}^{2 R} \right)'' {\rm e}^{2 A} + \left( {\rm e}^{2 R} \right)' \left( {\rm e}^{2 A} \right)'
+  \left( {\rm e}^{2 R} \right) \left( {\rm e}^{2 A} \right)''
+ 2 \, {\rm e}^{2R + 2 A} \, q_I \, N^{IJ} {\bar q}_J = 0 \;.
\label{eq:var-psi}
\end{eqnarray}
Now we use that on a solution to the first-order flow equations,
\begin{equation}
\left( {\rm e}^{2A} \right)'' 
= - 2 \, {\rm e}^{2A}  \, q_I \, N^{IJ} {\bar q}_J  \;.
\label{eq:rel-A-qq}
\end{equation}
This relation will be established below.
Then, using \eqref{eq:rel-A-qq} in 
\eqref{eq:var-psi} gives
\begin{eqnarray}
\left( {\rm e}^{2 R} \right)'' {\rm e}^{2 A} + \left( {\rm e}^{2 R} \right)' \left( {\rm e}^{2 A} \right)'
= \Big[\left( {\rm e}^{2 R} \right)' {\rm e}^{2 A} \Big]' = 0 \;,
\label{eq:cond-R}
\end{eqnarray}
which implies
\begin{equation}
\left( {\rm e}^{2 R} \right)' = C \, {\rm e}^{-2A} \;,
\label{eq:sol-R-1}
 \end{equation}
where $C$ denotes an integration constant. 
 
The variation of the 
action \eqref{eq:action-mu-def} with respect to $U$ also gives the equation 
\eqref{eq:cond-R}, while varying the action with respect to $R$ yields \eqref{eq:rel-A-qq},
which is satisfied on a solution to the first-order flow equations.

Finally, varying the action \eqref{eq:action-mu-def} with respect to $Y^I$ and using its flow equation in the form
\begin{equation}
N_{IJ} \, {\bar Y}''^J = \left( {\rm e}^A \, q_I  \right)' + i \, {\rm e}^{2A} \,
q_L \, N^{LP} \, 
F_{P JI} \, N^{JK} \, {\bar q}_K \;,
\label{eq:Y-2der}
\end{equation}
yields (upon integration by parts)
\begin{equation}
\left({\rm e}^{2R} \right)' \left({\bar Y}'^J - A' \, {\bar Y}^J \right) + {\rm e}^{2 R} \, 
{\bar Y}''^J = 0 
\label{eq:cond-Y2}
\end{equation}
 which, using \eqref{eq:sol-R-1}, can also be written as 
 \begin{equation}
\left({\rm e}^{2 R}\right)''  \, {\bar Y}^J + 2 \left({\rm e}^{2 R}\right)' \, {\bar Y}'^J+ 2 \, {\rm e}^{2 R}  \,  {\bar Y}''^J = 0\;.
\label{eq:2nd_order_eq_R}
\end{equation}

%%%%%%%%%%%%%%%%

Now, let us demonstrate the validity of \eqref{eq:rel-A-qq}.
This relation can be established by noticing that it equals
\begin{equation}
\left( {\rm e}^{2A} \right)''  + 2
N_{IJ} \, Y'^{I} {\bar Y}'^{\bar J} = 0
\label{eq:rel-A-Yd-Yd}
\end{equation}
on a solution to the first-order flow equations.  Using \eqref{eq:relAp-Y}, this in turn becomes
\begin{equation}
N_{IJ} \left( Y''^I \, {\bar Y}^J + Y^I \, {\bar Y}''^J \right) = i \left(F_{IJK} \, Y'^I \,
Y'^J \, {\bar Y}^K - {\bar F}_{\bar I \bar J \bar K} {\bar Y}'^I \, {\bar Y}'^J \, Y^K
\right) = 0\;.
\label{eq:cond-cons}
\end{equation}
The second equality in \eqref{eq:cond-cons} follows from
\begin{equation}
F_{IJK} \, Y'^I \, Y'^J = (F_{IJK} \, Y^I \, Y'^J)' - F_{IJK} \, Y^I \, Y''^J = 0\ ,
\label{eq:nicerelation}
\end{equation}
where \eqref{eq:homog-rel} was used in the last step. We now proceed to show the vanishing of
the left hand side of \eqref{eq:cond-cons}, which will then imply \eqref{eq:rel-A-Yd-Yd}.

Using \eqref{eq:nicerelation} and the first-order flow equation for $Y^I$,
we compute
\begin{equation}
Y''^I = N^{IK} \, \left({\rm e}^A \, {\bar q}_K  \right)' - i \, N^{IK}
{\bar F}_{\bar K \bar P \bar Q} \, Y'^P \, {\bar Y}'^Q \;.
\end{equation}
Inserting this into \eqref{eq:cond-cons} gives
\begin{equation}
\left({\rm e}^A \, q_I  \right)' \, Y^I  + \left({\rm e}^A \, {\bar q}_I  \right)' \, {\bar Y}^I = 0 \;.
\label{eq:qY-bar-qY}
\end{equation}
Next,  using \eqref{eq:q-Qhat-hhat}
we obtain
\begin{equation}
\left({\rm e}^A \, q_I  \right)' \, Y^I = \left( i \, \gamma - \psi \right)' \,
{\rm e}^A \, q_I \, Y^I - 2 i \, g \, A' \, {\rm e}^{2 A - \psi + i \, \gamma} \, \hat{h}_I \, Y^I \;.
\label{eq:const-cond}
\end{equation}
Using
\begin{equation}
A' = - {\rm e}^{-A}  q_I \, Y^I \;,
\end{equation}
which holds due to \eqref{eq:q-Y-real},
\eqref{eq:const-cond} becomes
\begin{equation}
\left({\rm e}^A \, q_I  \right)' \, Y^I = \left[
\left( i \, \gamma - \psi \right)' \,
-  2 i \, g  \, {\rm e}^{- \psi + i \, \gamma} \, W(Y)
\right] \, {\rm e}^A \, q_I \, Y^I
\;.
\label{eq:const-cond2}
\end{equation}
Inserting the flow equations for $\gamma'$
and $\psi'$ into this, and using the constraint \eqref{eq:cond-Z-W-Y}
yields
\begin{equation}
\left({\rm e}^A \, q_I  \right)' \, Y^I = 0 \;.
\end{equation}
Thus we conclude that \eqref{eq:qY-bar-qY} vanishes
on a solution to the first-order
flow equations, and hence also the left hand side of \eqref{eq:cond-cons}.

This concludes the discussion of equations resulting from the variation of the one-dimensional action 
\eqref{eq:action-mu-def}.  Next, we have to impose the Hamilton constraint \eqref{hamilton}. Inserting
the deformed line element \eqref{eq:line-def}, we find that the changes induced in \eqref{eq:outc-ham}
are proportional
to derivatives of $R$.  Thus, demanding $R = {\rm constant}$, we obtain that both the Hamilton constraint and
\eqref{eq:sol-R-1} are satisfied, provided
we set $C=0$
in the latter equation.
Finally, inserting $R = {\rm constant}$ in \eqref{eq:2nd_order_eq_R}
yields ${\bar Y}''^J = 0$.  Thus, we conclude that the class of non-extremal black brane solutions described
by the deformed line element \eqref{eq:line-def} with constant $R$ has a description in terms of the
first-order flow equations \eqref{eq:flow-U-psi-all} provided that the $Y^I$ are either constant or exhibit a growth that is linear in $r$.

Examples of non-extremal black brane solutions that fall into this class are 1) the non-extremal version
of the solution presented in \eqref{eq:exact-sol-gen}, which includes the solution discussed in \cite{Charmousis:2010zz},
2) the non-extremal deformation of the extremal domain wall solution \eqref{eq:attr-dw-int} and 3)
the non-extremal version of the interpolating solution near $AdS_4$ and near
$AdS_2\times \mathbb{R}^2$ discussed in section \ref{sec:interpol}.

\vskip 5mm

%%%%%%%%%%%%%%%%%%%%%%%%%%%%%%%%%%%%%%%%%%%%%%%%%%%%%%%%%%%%%%%%%
{\Large {\bf Acknowledgments}}

\vskip 3mm

We would like to thank Gianguido Dall'Agata, Stefanos Katmadas and Stefan Vandoren for valuable discussions.
SB, MH, SN and NO thank CAMGSD for hospitality. GLC thanks the Niels Bohr Institute and ASC-LMU for hospitality. This work is supported in part by the Excellence Cluster ``The Origin and the Structure of the Universe'' in Munich.
The work of GLC is supported by the Funda\c{c}\~ao para a Ci\^encia e a Tecnologia (FCT/Portugal). The work of MH and 
SN is supported by the German Research Foundation (DFG) within the Emmy-Noether-Program (grant number: HA 3448/3-1).
The work of NO is supported in part by the Danish National
Research Foundation project ``Black holes and their role in quantum gravity''.
The work of SB, GLC, MH and SN
is supported in part by the transnational cooperation FCT/DAAD grant ``Black Holes, duality and string theory".

%%%%%%%%%%%%%%%%%%%%%%%%%%%%%%%%%%%%%%

\pagebreak 

\begin{appendix}

\section{Special geometry}
\label{specialgeo}

As is well known, the Lagrangian describing the couplings of $N=2$ vector multiplets to $N=2$ supergravity
is encoded in a  holomorphic function $F(X)$, called the prepotential, that depends on  
$n + 1$ complex scalar fields $X^I$ ($I = 0, \dots, n$).  Here, $n$ counts the number of physical
scalar fields.  The 
coupling to supergravity requires $F(X)$ to be homogeneous of degree two, i.e. $F(\lambda X) = \lambda^2 F(X)$,
from which one derives the homogeneity properties
\begin{eqnarray}
&& F_I = F_{IJ} \, X^J \;, \nonumber\\
&& F_{IJK} \, X^K = 0 \;,
\label{eq:homog-rel}
\end{eqnarray}
where $F_I = \partial F(X)/\partial X^I \,,\,
F_{IJ} = \partial^2 F/\partial X^I \partial X^J$, etc. The $X^I$ are coordinates on the big moduli space,
while the physical scalar fields $z^i = X^i/X^0$ ($i = 1, \dots, n$) parametrize an $n$-dimensional complex
hypersurface, which is defined by the condition that the symplectic vector
$(X^I, F_I (X))$ satisfies the constraint
\begin{equation}
i \left( {\bar X}^I \, F_I -  {\bar F}_I \, X^I \right) = 1 \;.
\label{eq:einstein-norm}
\end{equation}
This can be written as
\begin{equation}
- N_{IJ} \, X^I \, {\bar X}^{J} = 1 \;,
\label{eq:constraint-sugra}
\end{equation}
where 
\begin{equation}
N_{IJ} = -i \left( F_{IJ} - {\bar F}_{IJ} \right) \;.
\label{eq:N-big}
\end{equation}
The constraint \eqref{eq:constraint-sugra} is solved by setting
\begin{equation}
X^I = {\rm e}^{K(z, {\bar z})/2} \, X^I (z) \;,
\end{equation}
where $K(z, {\bar z})$ is the K\"ahler potential, 
\begin{equation}
{\rm e}^{- K(z , \bar z) } = |X^0 (z)|^2 \, [ - N_{IJ} \, Z^I \, {\bar Z}^J ] 
\end{equation} 
with $Z^I (z) = (Z^0, Z^i) = (1, z^i)$.
Writing 
\begin{equation}
F(X) = \left(X^0\right)^2 \, {\cal F}(z) \;,
\end{equation}
which is possible in view of the homogeneity of $F(X)$, we obtain
\begin{equation}
F_{0} = X^0 \left(2 {\cal F}(z) - z^i \, {\cal F}_i \right) \;,
\end{equation}
where ${\cal F}_i = \partial {\cal F}/\partial z^i$.  In addition, we compute
\begin{eqnarray}
F_{00} &=& 2 {\cal F} - 2 z^i \, {\cal F}_i + z^i \, z^j \, {\cal F}_{ij} \;, \nonumber\\
F_{0j} &=& {\cal F}_j - z^i \, {\cal F}_{ij} \;, \nonumber\\
F_{ij} &=& {\cal F}_{ij} \;,
\label{eq:F-calF}
\end{eqnarray}
to obtain
\begin{equation}
 - N_{IJ} \, Z^I \, {\bar Z}^J = i \left[ 2\left( {\cal F} - {\bar {\cal F}} \right) - \left( z^i - {\bar z}^i \right)
 \left( {\cal F}_i + {\bar {\cal F}}_i \right) \right] \;, 
 \end{equation}
and hence
\begin{equation}
{\rm e}^{- K(z , \bar z) } = i \, |X^0 (z)|^2 \,
\left[ 2\left( {\cal F} - {\bar {\cal F}} \right) - \left( z^i - {\bar z}^i \right)
 \left( {\cal F}_i + {\bar {\cal F}}_i \right) \right] \;.
\label{eq:kahler-pot} 
 \end{equation}
The $X^I(z)$ are defined projectively, i.e. modulo multiplication by an arbitrary holomorphic function,
\begin{equation}
X^I(z) \rightarrow {\rm e}^{- f(z)} \, X^I (z) \;.
\label{eq:projectransf}
\end{equation}
This transformation induces the K\"ahler transformation
\begin{equation}
K \rightarrow K + f + \bar f 
\label{eq:kahlertransf}
\end{equation}
on the K\"ahler potential, while on the symplectic vector $(X^I, F_I (X))$
it acts as a phase transformation, i.e.
\begin{equation}
(X^I, F_I (X)) \rightarrow {\rm e}^{- \ft12 (f - \bar f)} \, (X^I, F_I (X)) \;.
\label{eq:U1transf}
\end{equation}
The resulting geometry for the space of physical scalar fields $z^i$ 
is a special K\"ahler geometry, with K\"ahler metric
\begin{equation}
g_{i  \bar{\jmath}} = \frac{\partial^2 K(z, \bar z)}{\partial z^i \, \partial {\bar z}^j} 
\label{eq:kaehler-metric}
\end{equation}
based on a K\"ahler potential of the special form \eqref{eq:kahler-pot}.

Let us relate the K\"ahler metric \eqref{eq:kaehler-metric} to the metric \eqref{eq:N-big} on the big moduli space.
Differentiating ${\rm e}^{-K}$ yields
\begin{eqnarray}
\partial_k {\rm e}^{- K} &=& - \partial_k K \, {\rm e}^{-K} =
 i \, |X^0(z)|^2 \, \left[ {\cal F}_k - {\bar {\cal F}}_k - 
(z^i - {\bar z}^i) {\cal F}_{ik} \right] + \partial_k \ln X^0(z) \, {\rm e}^{-K}
\;, \nonumber\\
\partial_k \partial_{\bar l} {\rm e}^{- K} &=& [- \partial_k \partial_{\bar l} K  +  
\partial_k K \, \partial_{\bar l} K ]\, {\rm e}^{-K} 
= [- g_{k \bar l} + 
\partial_k K \, \partial_{\bar l} K ]\, {\rm e}^{-K} \nonumber\\
&=& i \, |X^0 (z)|^2 \,
\left( {\cal F}_{kl} - {\bar {\cal F}}_{k l} \right) 
- \Big[\partial_k \ln X^0(z) \, \partial_{\bar l} \ln {\bar X}^0 (\bar z)
\nonumber\\
&& \qquad \qquad \qquad \qquad - \partial_k \ln X^0(z) \, \partial_{\bar l} K
- \partial_k K \, \partial_{\bar l} \ln {\bar X}^0 (\bar z) \Big] {\rm e}^{-K} \;.
\label{eq:ddK}
\end{eqnarray}
Using \eqref{eq:F-calF} we have
\begin{equation}
N_{ij} = - i \left({\cal F}_{ij} - {\bar {\cal F}}_{ij} \right) \;,
\end{equation}
and hence we infer from \eqref{eq:ddK} that
\begin{equation}
g_{i \bar{\jmath}} = N_{ij} \, |X^0|^2 + \frac{1}{|X^0 (z)|^2} {\cal D}_i X^0(z) {\cal D}_{\bar \jmath} \, {\bar X}^0 (\bar z) \;,
\label{eq:relgNX0}
\end{equation}
where
\begin{equation}
{\cal D}_i X^0 (z)= \partial_i X^0(z) + \partial_i K \, X^0(z)
\end{equation}
denotes the covariant derivative of $X^0(z)$ under the transformation \eqref{eq:projectransf}, i.e.
${\cal D}_i \left( {\rm e}^{-f} X^0(z) \right) = {\rm e}^{-f} \, {\cal D}_i X^0(z)$.

Next, let us consider the combination $N_{IJ} \, {\cal D}_{\mu} X^I \, {\cal D}^{\mu} {\bar X}^J$, where
the space-time covariant derivative ${\cal D}_{\mu}$ reads
\begin{equation}
{\cal D}_{\mu} X^I = \partial_{\mu} X^I + i {\cal A}_{\mu} X^I = 
\partial_{\mu} X^I + \frac12 \left(\partial_i K \, \partial_{\mu} z^i - \partial_{\bar \imath} K
\partial_{\mu} {\bar z}^i \right) X^I \;,
\label{eq:kah-conn}
\end{equation}
which is a covariant derivative for $U(1)$ transformations \eqref{eq:U1transf}.  The combination 
$N_{IJ} \, {\cal D}_{\mu} X^I \, {\cal D}^{\mu} {\bar X}^J$ is thus invariant under $U(1)$-transformations.
Observe that
\begin{equation}
{\cal D}_{\mu} X^0 = {\rm e}^{K/2} \, {\cal D}_i X^{(0)}(z) \, \partial_{\mu} z^i \;.
\label{eq:DDX0}
\end{equation}
Using \eqref{eq:constraint-sugra} we obtain
\begin{eqnarray}
N_{IJ} \, {\cal D}_{\mu} X^I \, {\cal D}^{\mu} {\bar X}^J &=& |X^0|^2 \, N_{ij} \, \partial_{\mu} z^i \,
\partial^{\mu} {\bar z}^j 
- \frac{1}{|X^0|^2} \, {\cal D}_{\mu} X^0 \,
{\cal D}^{\mu} \bar{X}^0 \nonumber\\
&&+ \frac{X^0}{{\bar X}^0} \, N_{i J} {\bar X}^J \, \partial_{\mu} z^i \, {\cal D}^{\mu} {\bar X}^0
+ \frac{{\bar X}^0}{X^0} \, N_{I j} X^I \, \partial_{\mu} {\bar z}^j \, {\cal D}^{\mu} X^0 \;.
\end{eqnarray}
Next, using
\begin{equation}
X^0 \, {\bar X}^J \, N_{k J} = \frac{1}{X^0(z)} \, {\cal D}_k X^0(z) \;,
\end{equation}
as well as \eqref{eq:relgNX0} and \eqref{eq:DDX0}
we establish
\begin{equation}
N_{IJ} \, {\cal D}_{\mu} X^I \, {\cal D}^{\mu} {\bar X}^J = g_{i \bar \jmath} \, \partial_{\mu} z^i \,
\partial^{\mu} {\bar z}^j \;,
\label{eq:relgN}
\end{equation}
which relates the kinetic term for the physical fields $z^i$ to the kinetic term for the fields $X^I$ on the big moduli space.  Observe that both sides of \eqref{eq:relgN} are invariant under K\"ahler transformations
\eqref{eq:kahlertransf}.

Using the relation \eqref{eq:relgN}, we express 
the bosonic Lagrangian (describing the coupling of $n$ vector multiplets to $N=2$ $U(1)$ gauged supergravity)
in terms of 
the fields $X^I$ of big moduli space,
\begin{equation}
L = \ft12  \mathfrak{R} - N_{IJ} \, {\cal D}_{\mu} X^I \, {\cal D}^{\mu} {\bar X}^J + \ft14 {\rm Im} {\cal N}_{IJ}
\, F_{\mu \nu}^I \, F^{\mu \nu J} - \ft14 {\rm Re} {\cal N}_{IJ}
\, F_{\mu \nu}^I \, {\tilde F}^{\mu \nu J} -  g^2 V(X, \bar X) \;,
\label{eq:lagN2}
\end{equation}
where
\begin{equation}
{\cal N}_{IJ} = {\bar F}_{IJ} + i \, \frac{N_{IK} \, X^K \, N_{JL} \, X^L}{X^M \, N_{MN} \, X^N} \;,
\end{equation}
which satisfies the relation
\begin{equation}
{\cal N}_{IJ} \, X^J = F_I \;.
\end{equation}
The flux potential reads
\begin{equation}
V = g^{i \bar \jmath} \, {\cal D}_i W \, \bar{\cal D}_{\bar \jmath} {\bar W} - 3 |W|^2 \;\;\;,\;\;\;
W =  h^I \, F_I  - h_I \, X^I \;,
\end{equation}
where ${\cal D}_i X^I = \partial_i X^I + \frac12 \partial_i K \, X^I$.
Here, $(h^I, h_I)$ denote the magnetic/electric fluxes. $AdS_4$ with cosmological constant
$\Lambda = - 3 g^2$ corresponds to a constant $W$ with
$|W| = 1$. Switching off the flux potential corresponds to setting $g=0$.

Using the identity (see (23) of \cite{deWit:1996ag})
\begin{equation}
N^{IJ} = g^{i \bar \jmath} \, {\cal D}_i X^I \, \bar{\cal D}_{\bar \jmath} {\bar X}^J  - X^I \, {\bar X}^J \;,
\label{eq:N-id-X}
\end{equation}
the flux potential can be expressed as
\begin{eqnarray}
V(X, \bar X) &=&  \left[g^{i \bar \jmath} \, {\cal D}_i X^I \, \bar{\cal D}_{\bar \jmath} {\bar X}^J - 3 X^I \bar{X}^J \right] 
\left(h^K \, F_{KI}  - h_I \right) \left( h^K \, {\bar F}_{KJ}  - h_J\right) \nonumber\\
&& = \left[N^{IJ} - 2 X^I \bar{X}^J \right]
 \, \left(h^K \, F_{KI}  - h_I \right) \left( h^K \, {\bar F}_{KJ}  - h_J\right) \nonumber\\
&& = N^{IJ} \, \partial_I \tilde{W} \partial_{\bar J} \bar{\tilde{W}} - 2 \, |\tilde{W}|^2 \;,
\label{eq:flux-pot}
\end{eqnarray}
where in the last line $\tilde{W}$ is expressed in terms of $U(1)$-invariant fields ${\tilde X}^I$, 
\begin{equation}
\tilde{W}= h^I \, F_I ({\tilde X}) - h_I \, {\tilde X}^I  = \left( h^I \, F_{IJ}  - h_J \right) \, {\tilde X}^J  
\;.
\label{eq:tilde-W}
\end{equation}

\section{First-order rewriting in minimal gauged supergravity}
\label{mingauge}
The  $ N\,=\,2$ Lagrangian in  minimal gauged supergravity is given in terms of the prepotential $F\,=\,-\,i\,(X^0)^2$. Only a single scalar field, $X^0$ is turned on, whose absolute value  is set to a constant, $|X^0|\,=\,\frac{1}{4}$ by the constraint (A.3), and the consequent bulk 1-D Lagrangian density for the  metric ansatz (2.1)  is given by 
\begin{equation}
{\cal L}_{1D}\,=\,{\rm e}^{2\,A\,+\,2\,U}\,[\,-\,A'^2\,-2\,A'\,U'\,+\,\frac{1}{4}\,{\rm e}^{-2\,U\,-\,4\,A}\,(\,|\hat{Q}_0|^2\,-\,3\,g^2\,|\hat{h}_0|^2 {\rm e}^{4\,A}\,)\,]\,.\,
\end{equation}
We now perform a first-order rewriting for this Lagrangian inspired by the corresponding rewriting for black holes in minimal gauged supergravity, as done in \cite{Lu:2003iv}. 
Denoting $\phi^1\,=\,A\,$, $\phi^2\,=\,U$ , $W\,=\,{\rm e}^U\,(\,|\hat{Q}_0|^2\,+\,g^2\,|\hat{h}_0|^2\,{\rm e}^{4\,A}\,+ \,\gamma\,{\rm e}^A\,)\,$,\footnote{Here $\gamma$ is an arbitrary real number.}
 and the matrix, $m\,=\,\begin{pmatrix}\,1 && 1\\ 1 && 0 \end{pmatrix}\,$, we can write the above Lagrangian density as 
\begin{equation}
{\cal L}_{1D}\,=\,-\,{\rm e}^{2\,A\,+\,2\,U}\,[\,{\phi^a}'\,m_{ab}\,{\phi^b}'\,\,+\,\frac{1}{4}\,{\rm e}^{-4\,U\,-\,4\,A}\,m^{ab}\,W_a\,W_b\,]\,.\,
\end{equation}
Here, $W_1\,=\,W_A$ is the derivative of $W$ w.r.t $A$ and $W_2\,=\,W_U\,$ is the derivative of $W$ w.r.t $U$, given explicitly as 
\begin{eqnarray}
W_A\,&=&\, \frac{{\rm e}^{2\,U}}{2\,W}\,(\,4\,g^2\, |\hat{h}_0|^2\,{\rm e}^{4\,A}\,+\,\gamma\,{\rm e}^A\,)\,,\,\nonumber \\
W_U\,&=&\,W\,.\,
\end{eqnarray}
The first-order rewriting is then simply,
\begin{equation}
{\cal L}_{1D}\,=\,-\,{\rm e}^{2\,A\,+\,2\,U}\,[m_{ab}\,(\,{\phi^a}'\,-\,\frac{1}{2}\,{\rm e}^{-\,2\,U\,-\,2\,A}\,m^{ac}\,W_c\,)\,(\,{\phi^b}'\,-\,\frac{1}{2}\,{\rm e}^{-\,2\,U\,-\,2\,A}\,m^{bd}\,W_d\,)\,]\,-\,{\phi^a}'\,W_a\,.\,
\end{equation}
Using the chain-rule of derivatives on $W\,=\,W\,(\,\phi^a\,)\,$, the last term in the above equation becomes ${\phi^a}'\,W_a\,=\,W'\,$. Hence the first-order rewritten Lagrangian can finally be written as 
\begin{equation}
{\cal L}_{1D}\,=\,-\,{\rm e}^{2\,A\,+\,2\,U}\,[m_{ab}\,(\,{\phi^a}'\,-\,\frac{1}{2}\,{\rm e}^{-\,2\,U\,-\,2\,A}\,m^{ac}\,W_c\,)\,(\,{\phi^b}'\,-\,\frac{1}{2}\,{\rm e}^{-\,2\,U\,-\,2\,A}\,m^{bd}\,W_d\,)\,]\,-\,W'\,.\,
\end{equation}
The first-order equations for the metric functions can be written as 
\begin{eqnarray}
A'\,&=&\,\frac{1}{2}\,{\rm e}^{-2\,U\,-\,2\,A}\,W_U\,,\,\\
U'\,&=&\,\frac{1}{2}\,{\rm e}^{-2\,U\,-\,2\,A}\,(\,W_A\,-\,W_U\,)\,\,.\,
\end{eqnarray}
The Hamiltonian density can be written as ${\cal H}\,=\,{\cal L}_{1D}\,+\,[\,{\rm e}^{2\,(\,A+\,U\,)}\,(\,2\,A'\,)\,]\,$.
On-shell, using the first order equations, this can be written as ${\cal H}\,=\,{\cal L}_{1D}\,+\,[\,{\rm e}^{2\,(\,A+\,U\,)}\,(\,W'\,)\,]\,$.
Substituting for ${\cal L}_{1D}$ from (B.6), we see that the on-shell Hamiltonian density becomes 
\begin{equation}
{\cal H}\,=\,-\,{\rm e}^{2\,A\,+\,2\,U}\,[m_{ab}\,(\,{\phi^a}'\,-\,\frac{1}{2}\,{\rm e}^{-\,2\,U\,-\,2\,A}\,m^{ac}\,W_c\,)\,(\,{\phi^b}'\,-\,\frac{1}{2}\,{\rm e}^{-\,2\,U\,-\,2\,A}\,m^{bd}\,W_d\,)\,]\,.\,
\end{equation}
This is trivially zero on-shell as the perfect squares vanish due to the first order equationss.
Hence the Hamiltonian constraint from General Relativity is satisfied for field configurations obeying these first-order equations. 
This completes the first-order rewriting for the Lagrangian in minimally gauged supergravity. All smooth black brane solutions to these first-order first order equationss are necessarily non-supersymmetric. The first non-supersymmetric black brane solutions in minimally gauged supergravity were written down in \cite{Chamblin:1999tk} where it was also shown that  the supersymmetric solutions are singular.
\par

There is a different rewriting of this Lagrangian which gives an $AdS_2\,\times\, {\mathbb R}^2$ background, given below as
\begin{eqnarray}
{\cal L}_{1D}\,&=&\,-\,[\,(\,(\,{\rm e}^A\,)^{'}\,)^2\,{\rm e}^{2\,U\,}\,+\,2\,{\rm e}^{A\,+\,U}\,(\,(\,{\rm e}^A\,)^{'}\,)\,(\,{\rm e}^{U^{'}}\,-\,\frac{1}{\sqrt{v_1}}\,)\,\nonumber\\
                      \,& &\,+\,2\,(\,(\,{\rm e}^A\,)^{'}\,)\,{\rm e}^{A\,+\,U}\,\frac{1}{\sqrt{v_1}}\,-\,\frac{1}{4}\,{\rm e}^{-\,2\,A}\,(\,|\hat{Q}_0|^2\,-\,3\,g^2\,|\hat{h}_0|^2 {\rm e}^{4\,A}\,)\,]\,.\,
\end{eqnarray}
The first-order first order equationss following from the rewriting above are  
\begin{eqnarray}
(\,{\rm e}^A\,)\,^{'}&=&\,0\,,\,\\
(\,{\rm e}^U\,)\,^{'}&=&\,\frac{r}{\sqrt{v_1}}\,,\,
\end{eqnarray}
with $v_1\,=\,2\frac{{\rm e}^{4\,A}}{|\hat{Q}_0|^2}\,$.
The Hamiltonian density vanishes on-shell provided ${\rm e}^{4\,A}\,=\,\frac{|\hat{Q}_0|^2}{3\,g^2\,|\hat{h}_0|^2}\,$. 
The $AdS_2 \,\times\,{\mathbb R}^2$ background which is the near-horizon black brane geometry in the presence of fluxes, is a solution to these first order equationss.

%%%%%%%%%%%%%%%%%%%%%%%%%%%%%%%%%%%%%%%%%%%%%%%%%%%%

\end{appendix}

%%%%%%%%%%%%%%%%%%%%%%%%%%%%%%%%%%%%%%%%%%%%%%%%%%%%%%%%%%%%%%%%%%

\addcontentsline{toc}{section}{References}
\providecommand{\href}[2]{#2}
\begingroup\raggedright\endgroup
\end{document}